\documentclass[11pt, a4paper]{article}
\usepackage{jheppub}

\usepackage{amsthm}
\usepackage{amsfonts}
\usepackage[active]{srcltx}

\usepackage{amsmath}
\usepackage{amssymb}
\usepackage{citesort}

\def\k #1{\mbox{\rsfs  #1}}

\def\XXint#1#2#3{{\setbox0=\hbox{$#1{#2#3}{\int}$ }
\vcenter{\hbox{$#2#3$ }}\kern-.65\wd0}}

\renewcommand{\theequation}{\arabic{section}.\arabic{equation}}

.tfm scaled 1000 
.tfm scaled 1000 
\font\rsfs=rsfs10.tfm scaled 1000 

\newcommand{\To}{\longrightarrow}

\title{Classical torus conformal block, ${\cal N}=2^{*}$ twisted superpotential
and the accessory parameter \\
of Lam\'{e} equation}

\author{Marcin Pi\c{a}tek}

\affiliation{Institute of Physics, University of Szczecin,\\
Wielkopolska 15, 70-451 Szczecin, Poland}

\affiliation{Bogoliubov Laboratory of Theoretical Physics,
Joint Institute for Nuclear Research,\\
Moscow Region, 141980 Dubna, Russia}

\emailAdd{piatek@fermi.fiz.univ.szczecin.pl}

\abstract{In this work the correspondence between the semiclassical limit of
the DOZZ quantum Liouville theory on the torus and the
Nekrasov--Shatashvili limit of the ${\cal N} = 2^*$
($\Omega$-deformed) ${\rm U(2)}$ super-Yang-Mills theory is used to
propose new formulae for the accessory parameter of the Lam\'{e} equation.
This quantity is  in particular crucial for solving the problem of
uniformization of the one-punctured torus.
The computation of the accessory parameters for torus and
sphere is an open longstanding problem which can however be solved if one succeeds to derive
an expression for the so-called classical Liouville action.
The method of calculation of the latter has been proposed
some time ago by Zamolodchikov brothers. Studying the semiclassical
limit of the four-point function of the quantum Liouville theory on the sphere they have derived the classical action for the
Riemann sphere with four punctures.
In the present work Zamolodchikovs idea is exploited in the case of the Liouville field theory on the torus.
It is found that the Lam\'{e} accessory parameter is
determined by the classical Liouville action on the one-punctured torus
or more concretely by the torus classical block evaluated on the saddle point
intermediate classical weight. Secondly, as an implication
of the aforementioned correspondence
it is obtained that the torus accessory parameter
is related to the sum of all rescaled column lengths of the so-called
``critical'' Young diagrams extremizing the instanton ``free energy''
for the ${\cal N} = 2^*$ gauge theory.
Finally, it is pointed out that thanks to the known relation the sum over the ``critical''
column lengths can be expressed in terms of a contour integral
in which the integrand is built out of certain special functions.}


\arxivnumber{1309.7672}

\begin{document}
\maketitle

\section{Introduction}
The name ``Lam\'{e} equation'' denotes in fact a class of related
ordinary second-order differential equations in the complex or
real domain which contain (explicitly or implicitly) certain
elliptic functions \cite{WW}. One of the most suitable 
forms of the Lam\'{e} equation in
practical applications  is the
so-called {\it Jacobian form}:
\begin{equation}\label{LameJacobi}
\frac{d^{2}\Psi}{d u^2}-\left[\,\kappa\, m\,{\rm sn}^2(u|m) + \mathrm{A}\,\right]\Psi\;=\;0.
\end{equation}
Eq.~(\ref{LameJacobi}) can be looked at as a one-dimensional Schr\"{o}dinger
equation:
$
-\Psi''(u)+V(u)\,\Psi(u) = {\cal E}\,\Psi(u)
$
with a {\it doubly periodic} potential $V(u)=\kappa\, m\,{\rm sn}^2(u|m)$
and the energy eigenvalue ${\cal E} =-\mathrm{A}$.
The potential is parameterized by the elliptic modular
parameter\footnote{We will use
two notations of Jacobi elliptic functions, i.e. with $m$:
${\rm sn}(u|m)$, ${\rm cn}(u|m)$, ${\rm dn}(u|m)$
and an alternative notation:
${\rm sn}(u,k)$, ${\rm cn}(u,k)$, ${\rm dn}(u,k)$
which uses a parameter $k=\sqrt{m}$.
For definition and properties of the Jacobi elliptic functions
see appendix \ref{ellipticf}.} $m$ of the
Jacobi ${\rm sn}$-function ${\rm sn}(u|m)$ and the constant $\kappa$.
In order to classify the solutions it is convenient to write $\kappa=\ell(\ell+1)$.
In particular, in the real domain if $\ell$ is a nonnegative integer
the energy spectrum consists of bands. There are $2\ell+1$ eigenfunctions
called {\it Lam\'{e} polynomials}\footnote{These are homogeneous polynomials of
degree $\ell$ in the elliptic functions:  ${\rm sn}$,  ${\rm cn}$,  ${\rm dn}$.}
associated with the boundaries of the energy gaps. For these
solutions the values of ${\cal E}$ are the solutions of a certain
algebraic equation \cite{WW,Maier}.\footnote{
Note, that for $\ell\in\mathbb{Z}^{+}$ the spectra of the Lam\'{e} and related finite-gap
periodic systems are characterized by a hidden bosonized nonlinear ${\cal N}=2$ supersymmetry, 
cf.~\cite{Ply1,Ply2,Ply3,Ply4,Ply5,Ply6,Ply7}.
It seems to be an interesting task to study whether this observation has something to do with
the so-called {\it Bethe/gauge} correspondence \cite{NekraSha,NekraSha2,NekraSha3}.}

Another important representation of the Lam\'{e} equation contains
the Weierstrass $\wp$-function:
\begin{equation}\label{LameW}
\frac{d^{2}\Psi}{d z^2}-\left[\,\kappa\,\wp(z) + \mathrm{B}\,\right]\Psi\;=\;0.
\end{equation}
Eq.~(\ref{LameW}) is known as the {\it Weierstrassian form} of the Lam\'{e} equation
and can be achieved from eq.~(\ref{LameJacobi}) by an appropriate change of the
independent variable (see appendix \ref{ellipticf}).
The {\it accessory parameters} $\mathrm{A}$ and $\mathrm{B}$
appearing in eqs.~(\ref{LameJacobi}) and (\ref{LameW}) are related to each other
in the following way\footnote{Recall, that the Weierstrass
$\wp$-function is doubly periodic on the complex plane with
periods $2\omega_1$, $2\omega_2$
and points $e_1, e_2, e_3$ are images $e_k = \wp(\omega_k)$
of the points $\omega_1$, $\omega_2$ and $\omega_3=-\omega_1-\omega_2$.}
\begin{equation}\label{AvsB}
\mathrm{B}\;=\; \mathrm{A}(e_1 - e_3) - \kappa\, e_3 \;\;\Leftrightarrow\;\;
\mathrm{A} \;=\;\frac{\mathrm{B}}{e_1 - e_3} - \frac{1}{3} \kappa (m+1).
\end{equation}

The substitution $\eta=\wp(z)$ converts eq.~(\ref{LameW}) to the third
version of the Lam\'{e} equation, commonly encountered in the literature,
the so-called {\it algebraic form}:
\begin{eqnarray}\label{algebraic}
\frac{d^{2}\Psi}{d \eta^2}+\frac{1}{2}\left[\frac{1}{\eta-e_1}
+\frac{1}{\eta-e_2}+\frac{1}{\eta-e_3}\right]\frac{d\Psi}{d \eta}
-\left[\frac{\kappa\,\eta + \mathrm{B}}{4(\eta-e_1)(\eta-e_2)(\eta-e_3)}
\right]\Psi\;=\;0.
\end{eqnarray}
Eq.~(\ref{algebraic}) ``lives'' on the Riemann sphere
$\mathbb{C}\cup\lbrace\infty\rbrace$. Its (regular) singular
points are located at $\eta=e_1, e_2, e_3, \infty$. The algebraic
form of the Lam\'{e} equation is the most appropriate form for
further generalizations, cf. \cite{Maier}.

Historically, the Lam\'{e} equation has first been obtained (by Lam\'{e})
by applying the method of separation of variables to the Laplace equation
in ellipsoidal coordinates \cite{WW}. More recently it has been noticed that
the Lam\'{e} equation arises in various physical contexts.
First, the Lam\'{e} potential can be considered as a good candidate
for a realistic model of a one-dimensional crystal \cite{Iachello}.
Other areas, where the Lam\'{e} equation is applicable, are
superconductivity \cite{Caputo}, certain version of the
Ginzburg-Landau theory \cite{Maier2}, and the cosmological models
\cite{Kantowski,Boyanovski,Greene,Kaiser,Ivanov}.\footnote{
For instance, in the theory of preheating the Lam\'{e} equation
has been recognized to determine quantum fluctuations of the inflaton field
\cite{Boyanovski,Greene,Kaiser,Ivanov}. Let us note,
that in several listed above applications the Lam\'{e} equation
plays the same role. Indeed,
regardless of the specific physical contexts one can observe
that the Lam\'{e} equation serves as the {\it stability equation}
or the {\it equation of small fluctuations}
around classical configurations associated with a certain class of basic potentials,
cf. \cite{M-K}.}
In mathematical physics applications the Lam\'{e} equation occurs in the
so-called Lie-algebraic approaches to the Schr\"{o}dinger equation \cite{Finkel}.

In pure mathematics the Lam\'{e} equation arises in the
uniformization theory of tori \cite{KRV}. More concretely, let $\mathfrak{T}$
denotes the Teichm\"{u}ller space for the one-punctured torus.
It is well known that $\mathfrak{T}$ has at least two distinct models.
The first one is the upper half plane
$\mathbb{U}=\left\lbrace\tau\in \mathbb{C}\,|\,{\rm Im}\,\tau>0\right\rbrace$. The
second model for $\mathfrak{T}$ is a subset of $\mathbb{R}^3$, namely
$
\mathbb{F} = \left\lbrace (x,y,z)\in\mathbb{R}^3\,
|\, x,y,z\; {\rm are}\, {\rm positive}
\,{\rm and}\; x^2+y^2+z^2=xyz\right\rbrace.
$
The question arises what is a relationship between $\mathbb{U}$ and $\mathbb{F}$.
It turns out that the mapping $\varphi:\mathbb{U}\to\mathbb{F}$ is determined by the
monodromy of the linearly independent solutions $(\Psi_1, \Psi_2)$ of the equation:
\begin{equation}\label{LameWII}
\frac{d^{2}\Psi}{d z^2}+\frac{1}{4}\left[\,\wp(z|{\cal L}) +
\mathrm{C}({\cal L})\,\right]\Psi\;=\;0,
\end{equation}
where $z\in \mathbb{C}-{\cal L}$ and ${\cal L}$ is a period lattice.
However, an explicit construction of $\varphi$ is difficult and still an open problem
since the accessory parameter $\mathrm{C}$ in the above equation is an undetermined constant,
cf. \cite{KRV}.

%

The Weierstrassian-form Lam\'{e} equation appears also in a
two-dimensional conformal field theory ($2d$ CFT) as a classical
limit of the null vector decoupling equation satisfied by
a torus two-point correlation function with a degenerate field.\footnote{
Surprisingly, it seems to be possible to obtain the
Lam\'{e} system (Hamiltonian and eigenfunctions)
from a certain matrix model.
Indeed, as has been observed in \cite{BMT}, in a suitable limit the so-called
loop equation of certain generalized matrix model yields
the equation closely related to the KZB equation \cite{Bernard,Etingof,FW}
(see also \cite{AT}).
The latter is known to reduces to Lam\'{e} equation in some cases.}
Moreover, one can observe that the accessory parameter $\mathrm{B}$ in this
equation is expressed in terms of the so-called {\it
classical Liouville action} on the one-punctured torus.
The latter quantity can be computed by means of
$2d$ CFT technics and recently discovered dualities, in particular
applying the correspondence between the classical limit of the quantum
DOZZ Liouville theory on the torus and the Nekrasov-Shatasvili
limit of the ${\cal N}=2^*$ ${\rm U}(2)$ super-Yang-Mills theory.

The aim of the present work is to find an analytical expression of
the Liouville classical action on the one-punctured torus
employing aforementioned technology and apply it to compute the
Lam\'e accessory parameter $\mathrm{B}$ ($\Leftrightarrow$
$\mathrm{A}$ and/or $\mathrm{C}$). The main motivation for this
line of research is the above mentioned monodromy problem for the
Lam\'e equation. Its solution is crucial not only for finding the
correspondence between models of the Teichm\"{u}ller space for
punctured torus but also for constructing a solution of the
Liouville equation on such surface, cf.
\cite{Menotti1,Menotti2,Menotti3,Menotti:2013bka}.

The organization of the paper is as follows.
In section 2 we briefly review interrelationships between Liouville theory
and the problem of computation of the accessory parameters of the Fuchsian
uniformization of the punctured Riemann sphere. The principal purpose of this
section is to recall the concept of the classical Liouville action and the \textit{classical
conformal block}  \cite{Zamolodchikov:1995aa}. For a long time the motivations
to study classical blocks were mainly confined to applications in pure mathematics,
in particular to the celebrated uniformization problem,
which roughly speaking is related to the construction of
conformal mappings between Riemann surfaces (RS)
admitting a simply connected universal covering and the three existing simply connected RS,
the sphere, the complex plane and the upper half plane.
The uniformization problem is well illustrated by the example of
the uniformization of the Riemann sphere with $n$ punctures.
Its uniformization may be associated to a Fuchsian equation whose form is
known up to some constants that are called accessory parameters.
Their computation is an open longstanding problem, which can however be solved	
if	we	succeed	to	derive	an analytical	expression	of	the	classical	block	
obtained	by performing the classical limit of the $n$-point correlation
function of the quantum Liouville field theory.
The importance of the classical blocks is not only limited to the uniformization theorem,
but gives also information about the solution of the Liouville equation on surfaces with punctures.
Recently, an interesting mathematical application of classical blocks emerged in the context
of Painlev\'{e} VI equation \cite{ZamClassBlock}.
Due to the recent discoveries the classical blocks are also relevant for physics,
since they are related to integrable models and to the instantonic sector of certain ${\cal N}=2$
supersymmetric gauge field theories \cite{NekraRosSha,Teschner,Piatek,FFPiatek}.
Moreover, lately classical conformal blocks have been of use to studies
of holographic principle and AdS/CFT correspondence \cite{Hartman}.

In section 3  we exploit the idea of brothers Zamolodchikov (see \cite{Zamolodchikov:1995aa})
in order to propose the form of the Liouville classical action $S_{\rm L}^{\rm torus}$
on the one-punctured torus. Our conjecture is that ($i$) $S_{\rm L}^{\rm torus}$ decomposes into
a sum of the three-point Liouville action on the sphere and the torus classical block; ($ii$)
factorization holds on the saddle point intermediate classical conformal weight.
Next, we consider the classical limit of the null vector
decoupling equation satisfied by
the torus two-point function with a degenerate field
and find an expression for the Lam\'{e} accessory parameter.
As has been already mentioned the latter is determined by the (one-punctured) torus classical action.

In section 4 we employ the AGT correspondence and express the toroidal classical block and then
the eigenvalue/accessory parameter
of  the Lam\'{e} equation in terms of the so-called effective
twisted superpotential of the ${\cal N}=2^*$ ${\rm U}(2)$ supersymmetric gauge theory.
In order to compute the latter quantity, i.e. the twisted superpotential, we use a straight-forward 
generalization of the calculation performed by Poghossian in \cite{Pogho1}.
We check that on the ``classical level'' the eigenvalue computed by means of the WKB method exactly
coincides with that obtained from the classical torus block. Finally, using a relation between the
instantonic sector of the twisted superpotential of the  ${\cal N}=2^*$ ${\rm U}(2)$ SYM theory
and the classical toroidal block we find that the Lam\'{e} accessory parameter is related
to the sum of column lengths of the so-called ``critical'' Young diagrams. It is shown that such sum can be
rewritten in terms of a contour integral in which the integrand is built out of certain special functions.

In section 5 we present our conclusions.
The problems that are still open and the possible extensions of the present work are discussed.

\section{Liouville theory and accessory parameters}
\subsection{Quantum and classical conformal blocks}
\label{QCcb}
Let $C_{g,n}$ denotes the Riemann surface with genus $g$ and $n$ punctures.
The basic objects of any two-dimensional conformal field theory
living on $C_{g}$ \cite{Belavin:1984vu,EO}
are the $n$-point correlation functions of primary vertex operators
defined on $C_{g,n}$.
Given a marking\footnote{A marking of the
Riemann surface $C_{g,n}$ (for definition see \cite{JT0}) is a pants decomposition of
$C_{g,n}$ together with the corresponding trivalent graph.}
$\sigma$ of the Riemann surface $C_{g,n}$
any correlation function can
be factorized according to the pattern given by a pants decomposition of
$C_{g,n}$ and written as a sum  (or an integral for theories with a continuous spectrum)
which includes the terms consisting of holomorphic and anti-holomorphic
conformal blocks times the three-point functions of the model
for each pair of pants. The Virasoro conformal block
${\cal F}_{c,\alpha}^{(\sigma)}[\beta]({\sf Z})$ on $C_{g,n}$, where
$\alpha\equiv(\alpha_1, \ldots,\alpha_{3g-3+n})$,
$\beta\equiv(\beta_1,\ldots,\beta_n)$
depends on the cross ratios of the vertex operators locations denoted symbolically
by ${\sf Z}$ and on the $3g-3+n$ intermediate conformal weights $\Delta_{\alpha_i}=
\alpha_i (Q -\alpha_i)$. Moreover, it depends on the $n$ external conformal weights
$\Delta_{\beta_a}=\beta_{a}(Q - \beta_a)$ and on the central charge $c$ which can be
parameterized as follows $c=1+6Q^2$ with $Q=b+b^{-1}$.

Conformal blocks are fully determined by the underlying conformal symmetry.
These functions possess an interesting, although not yet completely understood analytic
structure. In general, they can be expressed only as a formal power series and
no closed formula is known for its coefficients. Let us write down
two canonical examples which illustrate this fact.

Let $q=\textrm{e}^{2\pi i \tau}$  be the elliptic variable on the torus
with modular parameter $\tau$ then
the conformal block on $C_{1,1}$ is given by the following $q$-series:
\begin{eqnarray}
\label{torusblock} {\cal
F}_{c,\Delta}^{\tilde\Delta}(q)&=&q^{\Delta-\frac{c}{24}}
\left(1+\sum\limits_{n=1}^{\infty}{\cal
F}_{c,\Delta}^{\tilde\Delta,n}q^n \right),
\\
\label{torusCoeff} \mathcal{F}^{\tilde\Delta,
n}_{c,\Delta}&=&\sum\limits_{n=|I|=|J|}
\left\langle\nu_{\Delta,I},V_{\tilde\Delta}(1)\nu_{\Delta,J}\right\rangle
\;\Big[ G_{c,\Delta}\Big]^{IJ}.
\end{eqnarray}

Let $x$ be the modular parameter of the four-punctured sphere then
the $s$-channel conformal block on $C_{0,4}$ is defined as the following $x$-expansion:
\begin{eqnarray}
\label{block} {\cal
F}_{c,\Delta}\!\left[_{\Delta_{4}\;\Delta_{1}}^{\Delta_{3}\;\Delta_{2}}\right]\!(\,x)
&=& x^{\Delta-\Delta_{2}-\Delta_{1}}\left( 1 + \sum_{n=1}^\infty
{\cal
F}^{\,n}_{c,\Delta}\!\left[_{\Delta_{4}\;\Delta_{1}}^{\Delta_{3}\;\Delta_{2}}\right]
x^{n} \right),
\\\label{blockCeef}
{\cal F}^{\,n}_{c,\Delta}\!\left[_{\Delta_{4}\;\Delta_{1}}^{\Delta_{3}\;\Delta_{2}}\right]
&=&  \sum\limits_{n=|I|=|J|}
\left\langle\nu_{\Delta_4},V_{\Delta_3}(1)\nu_{\Delta,I}\right\rangle
\;
\Big[G_{c,\Delta}\Big]^{IJ}
\;
\left\langle\nu_{\Delta,J},V_{\Delta_2}(1)\nu_{\Delta_1}\right\rangle.
\end{eqnarray}

In the above equations
$\Big[ G_{c,\Delta}\Big]^{IJ}$ is the
inverse of the Gram matrix $ \Big[
G_{c,\Delta}\Big]_{IJ}=\langle \nu_{\Delta,I},\nu_{\Delta,J} \rangle
$ of the standard symmetric bilinear form in the Verma module ${\cal
V}_{\Delta}=\bigoplus_{n=0}^{\infty}{\cal V}_{\Delta}^{n}$,
\begin{eqnarray*}
{\cal V}_{\Delta}^{n}\;=\;\textrm{Span}\Big\lbrace
\nu^n_{\Delta,I}=L_{-I}\nu_{\Delta} = L_{-i_{k}}\ldots
L_{-i_{2}}L_{-i_{1}} \nu_\Delta
&:&\\
&&\hspace{-200pt} I=( i_{k}\geq \ldots\geq i_{1}\geq 1)
\;\textrm{an}\; \textrm{ordered} \;
\textrm{set}\; \textrm{of }\;\textrm{positive} \;\textrm{integers} \;\\
&&\hspace{-100pt} \textrm{of}\;
\textrm{the}\;\textrm{length}\;|I|\equiv i_{1}+\ldots+i_{k}=n
\Big\rbrace.
\end{eqnarray*}
The operator $V_{\Delta}$ in the matrix elements is the normalized
primary chiral vertex operator acting between the Verma modules
$$
\left\langle\nu_{\Delta_i},V_{\Delta_{j}}(z)\nu_{\Delta_k}\right\rangle
\;=\;z^{\Delta_i -\Delta_{j}-\Delta_k}.
$$
In order to calculate the matrix elements in
(\ref{torusCoeff}) and (\ref{blockCeef})
it is enough to know the covariance properties of the primary chiral
vertex operator with respect to the Virasoro algebra:
$$
\left[L_n , V_{\Delta}(z)\right] \;=\; z^{n}\left(z
\frac{d}{dz} + (n+1)\Delta
\right)V_{\Delta}(z)\,,\;\;\;\;\;\;\;\;\;\;n\in\mathbb{Z}.
$$
As the dimension of ${\cal V}^n_\Delta$ grows rapidly with $n$,
the calculations of conformal blocks coefficients by inverting
the Gram matrices become very laborious for higher orders.
A more efficient method based on recurrence relations for the coefficients
can be used \cite{Zam,Zamolodchikov:ie,FL,Pogho,HJStorus}.

Among  the issues concerning
conformal blocks which are still not fully understood there is the problem
of their {\it semiclassical limit}.
This is the limit in which all parameters of the conformal blocks
tend to infinity in such a way that their ratios are fixed.
It is commonly believed that such limit exists and the conformal blocks behave in this limit
exponentially with respect to ${\sf Z}$. This last property
can be heuristically justified in the case of conformal blocks on $C_{0,4}$ and $C_{1,1}$.

Indeed, the existence of the semiclassical limit of the Liouville
four-point correlation function with the projection on one
intermediate conformal family implies a
semiclassical limit of the quantum conformal block
with {\it heavy} weights $\Delta=b^{-2}\delta$,
$\Delta_i=b^{-2}\delta_i,$ with $\delta,\delta_i = {\cal O}(1)$ in the following form:
\begin{equation}
\label{defccb}
{\cal F}_{\!1+6Q^2,\Delta}
\!\left[_{\Delta_{4}\;\Delta_{1}}^{\Delta_{3}\;\Delta_{2}}\right]
\!(x)
\; \stackrel{b\,\to\,0}{\,\sim} \;
\exp \left\{
\frac{1}{b^2}\,f_{\delta}
\!\left[_{\delta_{4}\;\delta_{1}}^{\delta_{3}\;\delta_{2}}\right]
\!(x)
\right\}.
\end{equation}
The function
$f_{\delta}\!\left[_{\delta_{4}\;\delta_{1}}^{\delta_{3}\;\delta_{2}}\right]\!(\,x)$
is  called the classical conformal block \cite{Zamolodchikov:1995aa} or
with some abuse of terms, the ``classical action'' \cite{Zam0,Zam}.
The existence of the semiclassical limit (\ref{defccb}) has been postulated first
in \cite{Zam0,Zam} where it has been pointed out that the
classical block is related to a certain monodromy problem
of a null vector decoupling equation in a similar
way in which the classical Liouville action is related to the
Fuchsian uniformization. This relation has been further used to
derive the $\Delta\to \infty$ limit of the four-point conformal block
and its expansion in powers of the so-called elliptic variable.

Analogously, the existence of the semiclassical limit of the projected
Liouville torus one-point function implies that the semiclassical
limit of the torus one-point block  with heavy weights
$\Delta=b^{-2}\delta$, $\tilde\Delta=b^{-2}\tilde\delta$
with $\delta,\tilde\delta = {\cal O}(1)$ has the form:
\begin{equation}\label{classT}
{\cal F}^{\tilde\Delta}_{1+6Q^2,\Delta}(q)\;\stackrel{b\to 0}{\sim}\;
\exp\left\lbrace\frac{1}{b^2}f^{\tilde\delta}_{\delta}(q)\right\rbrace.
\end{equation}
The function $f^{\tilde\delta}_{\delta}(q)$ we shall call the classical
torus (or toroidal) conformal block.

It should be stressed once again that the exponential behavior
(\ref{defccb}) and/or (\ref{classT}) is a nontrivial
statement concerning the quantum conformal blocks.
Although there is no proof of this property,
it seems to be well confirmed together with its consequences by
sample numerical calculations \cite{HJP} and recent discoveries.

The classical conformal blocks are in general again available only
as power series with  coefficients calculated from the semiclassical
asymptotics and the power expansions of the quantum blocks.
The question arises how to sum up these series.
Surprisingly, one can find closed formulae for at least the four-point spherical
\cite{FFPiatek}
and the one-point toroidal classical blocks employing the AGT correspondence.

Indeed, a considerable progress in the theory of conformal
blocks and their applications has been achieved recently.
This is mainly due to the discovery of the Liouville/${\cal N}=2$ gauge
theories correspondence by Alday, Gaiotto and Tachikawa in 2009 \cite{AGT}.
The AGT conjecture states that the LFT correlators on the
Riemann surface $C_{g,n}$ with genus $g$ and $n$ punctures can be identified with the
partition functions of a class $T_{g,n}$ of four-dimensional ${\cal N}=2$
supersymmetric ${\rm SU}(2)$ quiver gauge theories.
A significant part of the AGT conjecture is an exact correspondence between the Virasoro
blocks on $C_{g,n}$ and the instanton sectors of the Nekrasov partition functions
of the gauge theories $T_{g,n}$. Very soon after its discovery,
the AGT hypothesis was extended to the SU(N)-gauge theories/conformal Toda correspondence
\cite{Wyllard,MMU(3),MMMM}.\footnote{Of course, there have been made
attempts to prove the AGT conjecture and its generalizations soon after its discovery.
Active studies of this duality have first led
to proofs of the AGT relations in certain special cases \cite{FL,MM2,Hadasz:2010xp}.
For more recent and more general achievements in this field, see
\cite{Alba:2010qc,Vartanov:2013ima,SV,MO,MCTan,Matsuo1,Matsuo2}.}

Let us recall that originally the Nekrasov partition functions
have been introduced to calculate the low energy effective ${\cal N}=2$
SUSY gauge theories prepotentials \cite{N,NekraOkun}.
The Seiberg-Witten prepotentials \cite{SW1,SW2} determine the low energy effective dynamics
of the four-dimensional ${\cal N}=2$ super-Yang-Mills theories and can be
recovered from the Nekrasov partition functions in the appropriate limit,
i.e. when the so-called $\Omega$-background parameters: $\epsilon_1$, $\epsilon_2$
appearing in the Nekrasov functions tend to zero.

The Nekrasov functions lead also to an interesting application when one of the
two $\Omega$-background parameters is non-zero. Such situation has been considered by
Nekrasov and Shatashvili in \cite{NekraSha}.
They observed that in the limit when one of the
$\Omega$-background parameters, say $\epsilon_2$, is zero and the second one,
i.e.~$\epsilon_1$, is kept finite ({\it Nekrasov-Shatashvili limit}) then one
can extract from the Nekrasov functions in this limit
the so-called {\it effective twisted superpotentials}.
These quantities determine the low energy effective dynamics of the
two-dimensional ($\Omega$-deformed) supersymmetric gauge theories.
Twisted superpotentials play also a prominent role in another context, namely in the so-called
{\it Bethe/gauge correspondence} \cite{NekraSha,NekraSha2,NekraSha3}
which maps supersymmetric vacua of the ${\cal N}=2$ two-dimensional
theories to Bethe states of quantum integrable systems (QIS's).
A result of that duality is that twisted superpotentials are identified
with {\it Yang's functionals} \cite{YY} which describe a spectrum of QIS's.

Looking at the AGT relations it is not difficult to realize that
the Nekrasov-Shatashvili
limit of the Nekrasov instanton partition functions corresponds
to the classical limit of conformal blocks.
Let us note that by combining the AGT duality and the Bethe/gauge
correspondence it is possible to link classical blocks to Yang's
functionals, cf. \cite{NekraRosSha,Teschner,Piatek}.

Using correspondence identifying classical blocks
and twisted superpotentials one can find a closed formulae for at least
spherical and toroidal  classical blocks.
As has been already mentioned, the relevant technical problem
of this strategy consists in the summation of the series defining
the classical block. This problem can be tackled on the gauge theory side
by means of the saddle point method \cite{N,NekraOkun,Pogho1,Fucito}.

\subsection{Classical Liouville action}
Let $C_{0,n}$ be the $n$-punctured Riemann sphere with complex coordinates chosen
in such a way that $z=\infty$. Consider the Liouville equation
\begin{equation}
\label{Liouville}
\partial_z\partial_{\bar z} \phi(z,\bar z) \;=\;
\frac{\varrho}{2}\, \textrm{e}^{\phi(z,\bar z)}
\end{equation}
with one of the following asymptotic behaviors
of the Liouville field $\phi(z,\bar z)$ near the punctures:
\begin{enumerate}
\item
case of {\it elliptic singularities}:
\begin{eqnarray}
\label{asymptot:elliptic}
&&\hspace{-50pt}\phi(z,\bar z) = \left\{
\begin{array}{lll}
-2\left(1-\xi_j \right)\log | z- z_j |  + O(1) & {\rm as } & z\to
z_j,
\hskip 5mm j = 1,\ldots,n-1,\\
-2\left(1+\xi_n \right)\log | z| + O(1) & {\rm as } & z\to \infty,
\end{array}
\right.
\\[3pt]
\label{Picard}
&&\hspace{10pt}
\xi_i \in \mathbb{R}_{>0}\;\;\;\textrm{for}\;\textrm{all}\;\;i=1,\ldots,n
\;\;\;\textrm{and}\;\;\sum\limits_{i=1}^{n}\xi_i < n-2;\nonumber
\end{eqnarray}
\item
case of {\it parabolic singularities} ($\xi_i \to 0$):
\begin{equation}
\label{asymptot:parabolic} \phi(z,\bar z) = \left\{
\begin{array}{lll}
-2\log |z- z_j |  -2\log \left|\log |z- z_j |\right| + O(1) & {\rm as } & z\to z_j, \\
-2\log |z| - 2\log \left|\log |z|\right| + O(1) & {\rm as } & z\to
\infty.
\end{array}
\right.
\end{equation}
\end{enumerate}
It is known that it exists a unique solution of eq.~(\ref{Liouville})
if one of the conditions
(\ref{asymptot:elliptic})
\cite{Picard1,Picard2,Troyanov}
or (\ref{asymptot:parabolic}) \cite{Heins} is satisfied.

One can define the Liouville action $S_{\rm L}[\phi]$ on $C_{0,n}$.
Because of the singular nature of the Liouville field at the punctures
such action has to be properly regularized:
\begin{eqnarray}
\label{action}
S_{\rm L}[\phi]
& = &
\frac{1}{4\pi}
\lim_{\epsilon\to 0}
S_{\rm L}^\epsilon[\phi],
\\
S_{\rm L}^\epsilon[\phi]
& = &
\int\limits_{X_\epsilon}\!d^2z
\left[\left|\partial\phi\right|^2 + \varrho\,{\rm e}^{\phi}\right]
+  \sum\limits_{j=1}^{n-1}\left(1-\xi_j\right)
\hspace{-4mm}\int\limits_{|z-z_j|=\epsilon}\hspace{-4mm}|dz|\ \kappa_z \phi
+\left(1+\xi_n\right)
\hspace{-2mm}\int\limits_{|z|=\frac{1}{\epsilon}}\hspace{-2mm}|dz|\ \kappa_z
\phi\nonumber
\\
&&
- 2\pi\sum\limits_{j=1}^{n-1}\left(1-\xi_j\right)^2\log\epsilon
- 2\pi\left(1+\xi_n\right)^2\log\epsilon,\label{regLFTaction}
\end{eqnarray}
\( X_\epsilon = {\mathbb C}\setminus\left\{\left(\bigcup_{j=1}^n |z-z_j|< \epsilon\right)
\cup \left(|z|>\frac{1}{\epsilon}\right)\right\}. \) The prescription
given in
eqs.~(\ref{action}) and (\ref{regLFTaction})
is valid for parabolic singularities (corresponding to
$\xi_j = 0$) as well.

It is well known mathematical fact that the critical value $S_{\rm L}^{\,\rm
cl}[\phi]$ of the Liouville action functional $S_{\rm L}[\phi]$
on $C_{0,n}$ (the classical Liouville action)
is the generating function for the accessory parameters $c_j$ of the
Fuchsian uniformization of the punctured Riemann sphere, i.e.:
\begin{equation}
\label{PC} c_j \;=\; -\frac{\partial S_{\rm L}^{\,\rm
cl}[\phi]}{\partial z_j}.
\end{equation}
Primarily, this formula has been derived within the so-called
{\it geometric path integral approach} to the
quantum Liouville theory by analyzing the quasi-classical limit of the
conformal Ward identity \cite{Tak}.
Then, for parabolic singularities formula (\ref{PC}) has been proved by Takhtajan and Zograf.
The details can be found in \cite{TZ2}. In ref.~\cite{ZoTa2} the extension of \cite{TZ2} to
compact Riemann surfaces has been presented. For general elliptic singularities eq.~(\ref{PC})
has been proved in \cite{TZ} and non rigorously derived in \cite{Cantini:2001wr}.
It is also possible to construct the Liouville action functional satisfying
(\ref{PC}) for the so-called hyperbolic singularities (holes)
on the Riemann sphere, see \cite{Hadasz:2003kp}.

On the other hand one can observe that in the case of the
four-punctured sphere with singularities located at
$z_4 = \infty, z_3 = 1, z_2 = x, z_1 = 0$
the Fuchsian differential equation
$$
\label{Fuchs2}
\partial^{2}_{z}\Psi(z)+\left[\frac{\delta_1}{z^2}
+\frac{\delta_2}{(z-x)^2} +\frac{\delta_3}{(1-z)^2}
+\frac{\delta_1+\delta_2+\delta_3-\delta_4}{z(1-z)}
+\frac{x(1-x)c_{2}(x)}{z(z-x)(1-z)}\right]\Psi(z)=0
$$
with an accessory parameter $c_{2}(x)$
given by the derivative w.r.t. $x$ of the four-point classical action
can be obtained from the classical limit $b\to 0$ of certain null vector
decoupling equation. Concretely, from the equation
\begin{eqnarray*}
\label{Fuchs1} && \left[ \frac{\partial^2}{\partial z^2}
-b^2\left(\frac{1}{z} - \frac{1}{1-z}\right)\frac{\partial}{\partial
z} \right] G(z,x)
=
\\
\nonumber &&\hspace{-20pt}-b^2 \left[\frac{\Delta_1}{z^2} +
\frac{\Delta_2}{(z-x)^2} +  \frac{\Delta_3}{(1-z)^2} +
\frac{\Delta_1 \!+\! \Delta_2 \!+\!\Delta_3
\!+\!\Delta_{-\frac{b}{2}}\!-\!\Delta_4}{z(1-z)} + \frac{x(1-x)}{z(z-x)(1-z)}
\frac{\partial}{\partial x}\right]G(z,x)
\end{eqnarray*}
satisfied by the five-point function
$$
G(z,x) \; \equiv \; \Big\langle {\sf V}_4(\infty,\infty){\sf V}_3(1,1)
{\sf V}_{-\frac{b}{2}}(z,\bar z){\sf V}_2(x,\bar
x){\sf V}_1(0,0) \Big\rangle
$$
with a degenerate field
$$
{\sf V}_{\alpha=-\frac{b}{2}},
\;\;\;\;\;\;\;\;\;\;\;
\Delta_{-\frac{b}{2}}\;=\;\Delta_{\alpha=-\frac{b}{2}}
\;=\;\alpha(Q-\alpha) \;=\;-\frac{1}{2}-\frac{3}{4}\,b^2,
\;\;\;\;\;\;Q\;=\;b+\frac{1}{b}
$$
and four heavy primary operators
${\sf V}_{\Delta_i}$, $\Delta_i=b^{-2}\,\delta_i$,
$\delta_i = {\cal O}(1)$.

Analogously, the Weierstrassian-form
Lam\'{e} equation with the accessory parameter
determined by the torus classical one-point action can be recovered
from the null vector  decoupling equation satisfied by the
torus two-point correlation function with one degenerate field
(see subsection \ref{nvd}).

\subsection{Zamolodchikov's conjecture}
Hence, one can compute
the accessory parameters once the classical action is known.
The latter can be derived
by performing the classical limit of the DOZZ quantum LFT
correlation functions. In particular, Zamolodchikov brothers
\cite{Zamolodchikov:1995aa}
studying the classical limit of the four-point function
of the quantum Liouville theory on the sphere argued that
the classical Liouville action  with four elliptic/parabolic
singularities located at $z_4 = \infty, z_3 = 1, z_2 = x, z_1 = 0$
can be expressed as follows:
\begin{eqnarray}
\label{clasfact}
&&\hspace{-30pt} S^{\rm cl}_{\rm L}(\delta_4,\delta_3,\delta_2,\delta_1;x) =\\
&& S^{\rm cl}_{\rm L}(\delta_4,\delta_3,\delta_s(x))
+
S^{\rm cl}_{\rm L}(\delta_s(x),\delta_2,\delta_1)
-\,f_{\delta_s(x)}
\!\left[_{\delta_{4}\;\delta_{1}}^{\delta_{3}\;\delta_{2}}\right](x)
-\bar f_{\delta_s(x)}
\!\left[_{\delta_{4}\;\delta_{1}}^{\delta_{3}\;\delta_{2}}\right](\bar x).
\nonumber
\end{eqnarray}
Indeed, the four-point function of the DOZZ theory can be  defined
as an integral of $s$-channel conformal blocks and DOZZ couplings over
the continuous spectrum of the theory.
In the semiclassical limit $b\to 0$ 
the integrand can be expressed
in terms of three-point classical Liouville actions and the classical block,
and the integral itself is dominated by the saddle point $\Delta_s=\frac{1}{b^2}\,\delta_s(x)$.
One thus gets the factorization (\ref{clasfact}).

Concluding, in order to construct the four-point classical action
via Zamolodchikov's prescription one needs the following data:
\begin{itemize}
\item[(a)] the classical three-point Liouville action
$S^{\rm cl}_{\rm L}(\delta_3,\delta_2,\delta_1)$ for
the classical weights $\delta_1,\delta_2,\delta_3$ at the locations $0,1,\infty$;
\item[(b)] the four-point
classical conformal block on the sphere $f_{\delta}
\!\left[_{\delta_{4}\;\delta_{1}}^{\delta_{3}\;\delta_{2}}\right]\!(x)$ ;
\item[(c)]  the $s$-channel saddle point
conformal weight $\delta_s(x) = \frac{1}{4}+p_{s}^{2}(x)$ where
the  $s$-channel saddle point momentum $p_{s}(x)$ is
determined by the saddle point condition ($p\in \mathbb{R}$):
\begin{eqnarray}
\Big({\partial \over \partial p}\,S_{\rm L}^{\rm
cl}(\delta_4,\delta_3,{\textstyle {1\over 4}} +p^2)
+
{\partial \over \partial p} S_{\rm L}^{\rm
cl}({\textstyle {1\over 4}} +p^2,\delta_2,\delta_1)\nonumber
-
2{\rm Re}\,{\partial \over \partial p}\,f_{{\textstyle {1\over 4}} +p^2}
\!\left[_{\delta_{4}\;\delta_{1}}^{\delta_{3}\;\delta_{2}}\right](x)
\Big)\Big|_{p=p_s}\;=\;0.\nonumber
\end{eqnarray}
\end{itemize}

Let us stress that we will exploit the above idea in the present work
in order to compute the classical action on the one-punctured torus.

The semiclassical limit should be independent of the choice of the channel in
the representation of the DOZZ four-point function.
Therefore, one gets the consistency conditions
known as the {\it classical bootstrap equations} \cite{HJP}:
\begin{eqnarray}
\nonumber
&& \hspace{-80pt} S^{\rm cl}_{\rm L}(\delta_4,\delta_3,\delta_s(x))
+
S^{\rm cl}_{\rm L}(\delta_s(x),\delta_2,\delta_1)
-\,f_{\delta_s(x)}
\!\left[_{\delta_{4}\;\delta_{1}}^{\delta_{3}\;\delta_{2}}\right](x)
-\bar f_{\delta_s(x)}
\!\left[_{\delta_{4}\;\delta_{1}}^{\delta_{3}\;\delta_{2}}\right](\bar x)
\\[10pt]
\label{clasboot}
&=&
S^{\rm cl}_{\rm L}(\delta_4,\delta_1,\delta_t(x))
+
S^{\rm cl}_{\rm L}(\delta_t(x),\delta_2,\delta_3)\\
\nonumber
&&
\hspace{70pt}-\,f_{\delta_t(x)}
\!\left[_{\delta_{4}\;\delta_{3}}^{\delta_{1}\;\delta_{2}}\right](1-x)
-\bar f_{\delta_t(x)}
\!\left[_{\delta_{4}\;\delta_{3}}^{\delta_{1}\;\delta_{2}}\right](1-\bar x)
\\[10pt]
\nonumber
&=&  2\delta_2 \log x\bar x + S^{\rm cl}_{\rm L}(\delta_1,\delta_3,\delta_u(x))
+
S^{\rm cl}_{\rm L}(\delta_u(x),\delta_2,\delta_4)\\
\nonumber
&&
\hspace{70pt}-\,f_{\delta_u(x)}
\!\left[_{\delta_{1}\;\delta_{4}}^{\delta_{3}\;\delta_{2}}\right]\left({1\over x}\right)
-\bar f_{\delta_u(x)}
\!\left[_{\delta_{1}\;\delta_{4}}^{\delta_{3}\;\delta_{2}}\right]\left({1\over \bar x}\right).
\end{eqnarray}
The saddle weights $\delta_t(x), \delta_u(x)$ in the $t$- and $u$-channel  are simply related to
the $s$-channel saddle point classical weight:
$$
\delta_t(x) = \delta_s(1-x),\;\;\;\;\;\;\delta_u(x) = \delta_s\left({1\over x}\right).
$$

There is a nice geometric interpretation of the saddle point conformal weight
$\delta_i(x)$. Let us recall that the classical solution describes a unique hyperbolic geometry
with singularities at the locations of conformal weights. For
 elliptic, parabolic and hyperbolic weights one gets
conical singularities, punctures and holes with geodesic boundaries
respectively \cite{sei,Hadasz:2003he,Hadasz:2003kp}.
In the latter case the classical conformal weight $\delta$ is related to the length
$\ell$ of the corresponding hole by
\begin{equation}
\label{glength} \delta =\frac{1}{4}+
\frac{\varrho}{4}\left(\frac{\ell}{2\pi}\right)^2,
\end{equation}
where the scale of the classical configuration is set by the condition $R=-{\varrho/2}$
imposed on the constant scalar curvature $R$.

In the case of four singularities at the standard locations
$0,x,1,\infty$ there are  three closed ge\-o\-de\-sics $\Gamma_s,
\Gamma_t, \Gamma_u$ separating the singular points into pairs
$(x,0|1,\infty),$ $(x,1|0,\infty)$ and $(x,\infty|0,1)$
respectively. Since the spectrum of DOZZ theory is hyperbolic the
singularities corresponding to the saddle point weights
$\delta_i(x)$ are geodesic holes. One may expect that these
weights are related to the lengths $\ell_i$ of the closed
geodesics $\Gamma_i$ in corresponding channels \cite{HJP}:
\begin{equation}
\label{glengthII} \delta_i(x) =\frac{1}{4}+
\frac{\varrho}{4}\left(\frac{\ell_i(x)}{2\pi}\right)^2,\;\;\;i=s,t,u.
\end{equation}

\section{Lam\'{e} accessory parameter from Liouville theory}
\subsection{Semiclassical Liouville one-point function on the torus}
Let $\tau$ be the torus modular parameter and
$q={\rm e}^{2\pi i \tau}$, $\bar q={\rm e}^{-2\pi i \bar\tau}$.
The Liouville one-point function on the torus (expressed
in terms of $2d$ CFT quantities defined on the complex plane) reads as follows
\begin{eqnarray}
\label{L1pointA}
\left\langle\,{\sf V}_{\!\beta}(1)\,\right\rangle_{\tau}&\equiv &
{\rm Tr}_{\cal H}\left(q^{L_0 - c/24} {\bar q}^{{\bar L}_0 - c/24}
\,{\sf V}_{\!\beta}(z,\bar z)\right)\Big|_{z=1}
\\
&=&
\int\limits_{\mathbb{R}^{+}}d{\rm P}\, C\!\left(\bar\alpha_{\rm P},\beta, \alpha_{\rm P}\right)
\left|{\cal F}^{\Delta_\beta}_{c,\Delta_{\alpha_{\rm P}}}(q)\right|^2,
\label{L1pointB}
\end{eqnarray}
where
\begin{eqnarray}
\alpha_{\rm P} &=& \frac{Q}{2} + i{\rm P},
\;\;\;\;\;\;
\bar\alpha_{\rm P} \;=\; \frac{Q}{2} - i{\rm P} \;=\; Q-\alpha_{\rm P},
\;\;\;\;\;\;
{\rm P}\in\mathbb{R}^{+};
\\
\label{be}
\beta &=& \frac{Q}{2}(1+\xi),
\;\;\;\;\;\;
\xi\in[0,1)
\end{eqnarray}
and
\begin{eqnarray}
\label{para2}
\Delta_{\alpha_{\rm P}} &=& \alpha_{\rm P}(Q-\alpha_{\rm P})\;=\;
\frac{Q^2}{4}+{\rm P}^{2}\;\equiv\;\Delta(\rm P)\equiv\Delta,
\\
\label{para1}
\Delta_\beta &=& \beta(Q-\beta),
\\
\label{para3}
c &=& 1+6Q^2,
\;\;\;\;\;\;\;\;
Q\;=\;b+b^{-1}.
\end{eqnarray}
The trace in (\ref{L1pointA}) is taken over the basis of
the Liouville Hilbert space \cite{JT1}:
$$
{\cal H}\;=\;\int\limits_{\mathbb{R}^{+}}^{\oplus}d{\rm P}\;
{\cal V}_{\Delta({\rm P})}\otimes{\cal V}_{\Delta({\rm P})}.
$$
The operator ${\sf V}_{\!\beta}$ in (\ref{L1pointA}) is the primary Liouville
vertex operator with the conformal weight $\Delta_\beta$.
It has been assumed that ${\sf V}_{\!\beta}$ is a heavy field
($\Delta_\beta\stackrel{b\to 0}{\sim} b^{-2}\cdot{\rm const.}$
$\Leftrightarrow$ $\beta\stackrel{b\to 0}{\sim}b^{-1}\cdot{\rm const.}$).
Moreover, the operator ${\sf V}_{\!\beta}$ corresponds to
the so-called elliptic or parabolic ($\xi=0$) singularity (cf. condition (\ref{be})).
The integrand in (\ref{L1pointB}) is built out of the DOZZ structure constant
\cite{Zamolodchikov:1995aa,DO}:
\begin{eqnarray*}
\label{DOZZ}
&&C(\alpha_1, \alpha_2, \alpha_3)\;=\; \left[
\pi\mu\gamma(b^2)b^{2-2b^2}\right]^{(Q-\alpha_1 - \alpha_2 -
\alpha_3)/b}
\times\nonumber
\\[5pt]
&&\hspace{-50pt}\frac{\Upsilon_{0}\Upsilon(2\alpha_1)\Upsilon(2\alpha_2)\Upsilon(2\alpha_3)}
{\Upsilon(\sum \alpha_i - Q)\Upsilon(Q+\alpha_1 - \alpha_2 -
\alpha_3) \Upsilon(\alpha_1 + \alpha_2 - \alpha_3)\Upsilon(\alpha_1
- \alpha_2 + \alpha_3)},
\end{eqnarray*}
$$
\gamma(x)\equiv\frac{\Gamma(x)}{\Gamma(1-x)},
\;\;\;\;\;\;\;\;\;\;
\Upsilon_{0}=\frac{d\Upsilon(x)}{d x}\Big|_{x=0}
$$
and the torus one-point conformal block ${\cal F}^{\Delta_\beta}_{c,\Delta}(q)$
defined in (\ref{torusblock})--(\ref{torusCoeff}).

Now, we want to find the limit $b\to 0$ of the
one-point function (\ref{L1pointA})--(\ref{L1pointB}) in the case when all the
conformal weights $\Delta_{\alpha_{\rm P}}$, $\Delta_\beta$ are heavy
\cite{Zamolodchikov:1995aa}, i.e.:
\begin{equation}\label{heavyW}
\Delta_{{\alpha_{\rm P}}}\equiv\Delta\;\stackrel{b\to 0}{\sim}\;\frac{1}{b^2}\,\delta,
\;\;\;\;\;\;\;\;
\Delta_{\beta}\;\stackrel{b\to 0}{\sim}\;\frac{1}{b^2}\,\tilde\delta,
\;\;\;\;\;\;\;\;
\delta, \tilde\delta \;=\; {\cal O}(1).
\end{equation}
As has been already assumed the external weight $\Delta_{\beta}$ is heavy,
\begin{equation}\label{betaaa}
\beta\;\stackrel{b\to 0}{\sim}\;\frac{1}{2b}(1+\xi).
\end{equation}
The corresponding classical conformal weight $\tilde\delta$ is defined as follows
\begin{equation}
\label{betaa}
\tilde\delta\;=\;\lim\limits_{b\to 0} b^2 \Delta_{\beta}
= \frac{1}{4}\left(1-\xi^2\right),
\;\;\;\;\;\;\;\;\xi\in[0,1)
\;\;\;\;\Leftrightarrow\;\;\;\;
0<\tilde\delta\leq\frac{1}{4}.
\end{equation}
In the case of the intermediate weight $\Delta_{{\alpha_{\rm P}}}$, if we
rescale the integration variable ${\rm P}=\frac{p}{b}$, then
\begin{equation}
\label{alphaa}
\alpha_{\rm P}=\frac{Q}{2}+i\frac{p}{b}\;\stackrel{b\to 0}{\sim}\;
\frac{1}{2b}(1+2ip) \;\;\Leftrightarrow\;\;
\delta=\lim\limits_{b\to 0} b^2 \Delta_{\alpha_{\rm P}} = \frac{1}{4}+p^2,
\;\;\;p\in\mathbb{R}^{+}.
\end{equation}

Let us recall that
the elliptic/parabolic ($\xi=0$) classical weights $0<\tilde\delta\leq\frac{1}{4}$
are related to the parabolic/elliptic singularities,
i.e. conical singularities with an opening angle $2\pi\xi$.
The hyperbolic weights $\delta > \frac{1}{4}$ correspond to the hyperbolic singularities
--- holes with geodesic boundary (as has been already mentioned
the classical hyperbolic weight is related to the length of the corresponding
hole).

Let us turn to the problem of finding the $b\to 0$ limit
of the one-point function (\ref{L1pointA2})--(\ref{L1pointB2}).
First let us determine the asymptotical behavior of the integrand in
(\ref{L1pointB}) when $b\to 0$.

In \cite{Hadasz:2003he} it has been found that for the hyperbolic spectrum:
\begin{equation}\label{alphas}
\alpha_j \;=\; \frac{Q}{2}(1+i\lambda_j)
\;\stackrel{b\to 0}{\sim}\;
\frac{1}{2b}(1+i\lambda_j),
\;\;\;\;\;\; \lambda_j \in \mathbb{R},
\;\;\;\;\;
j=1,2,3
\end{equation}
the DOZZ three-point function in the limit $b\to 0$ behaves as follows
\begin{eqnarray}\label{DOZZlimit}
C\!\left(\alpha_1, \alpha_2, \alpha_3\right) &\sim & \exp\left\lbrace -\frac{1}{b^2}
\left[\sum\limits_{\sigma_1, \sigma_2 = \pm}
F\left( \frac{1+i\lambda_1}{2} + \sigma_1 \frac{i\lambda_2}{2} + \sigma_2 \frac{i\lambda_3}{2}\right)\right.\right.
\nonumber\\
&&+\sum\limits_{j=1}^{3}\left(H(i\lambda_j)+\frac{1}{2}\pi |\lambda_j|\right)+\frac{1}{2}\log(\pi\mu b^2)
\nonumber\\
&&\left.\left.-i\sum\limits_{j=1}^{3}\lambda_j \left( 1-\log|\lambda_j|+\frac{1}{2}\log(\pi\mu b^2)\right)
+ {\rm const.}
\right]
\right\rbrace,
\end{eqnarray}
where
$$
F(x)\;=\;\int\limits_{\frac{1}{2}}^{x}dy \log\frac{\Gamma(y)}{\Gamma(1-y)},
\;\;\;\;\;\;\;\;\;\;\;
H(x)\;=\;\int\limits_{0}^{x}dy \log\frac{\Gamma(-y)}{\Gamma(y)}.
$$
At this point, a few comments are in order.
The expression in the square brackets should correspond to the known expression
for the classical Liouville action $S_{\rm L}^{(3)}[\varphi]$
on the sphere with three hyperbolic singularities (holes).
Such classical action has been constructed in \cite{Hadasz:2003he}
(see also \cite{Hadasz:2003kp}).
The construction of $S_{\rm L}^{(3)}[\varphi]$
relies on a solution of a certain monodromy problem
for the Fuchsian differential equation:
$$
\frac{d^2 \Phi}{dz^2} + \sum_{k=1}^{n}\left[\frac{\delta_k}{(z - z_k)^2}+
\frac{c_k}{z - z_k}\right]\Phi \;=\; 0
$$
with hyperbolic singularities ($\delta_k$'s are hyperbolic).
In this way one can find the form of the $n$-point classical action
$S_{\rm L}^{(n)}[\varphi]$
up to of at most $n-3$ undetermined constants
$c_k$. $S_{\rm L}^{(n)}[\varphi]$
satisfies Polyakov's formula \cite{Hadasz:2003kp}:
\begin{equation}\label{Polyakov}
\frac{\partial}{\partial z_j}S_{\rm L}^{(n)}[\varphi] \;=\; -c_j.
\end{equation}
In the case when $n=3$ the {\it Fuchsian accessory parameters} $c_k$ are known
and the classical action can be determined from eq. (\ref{Polyakov}).
For the standard locations of singularities $z_1 = 0$, $z_2 = 1$, $z_3 = \infty$
this yields \cite{Hadasz:2003he}:
\begin{eqnarray}\label{cl}
Q^2 S_{\rm L}^{(3)}[\varphi] &=&
Q^2 \left[ \sum\limits_{\sigma_1, \sigma_2 = \pm}
F\left( \frac{1+i\lambda_1}{2} + \sigma_1 \frac{i\lambda_2}{2} + \sigma_2 \frac{i\lambda_3}{2}\right)\right.
\nonumber\\
&+&\left.\sum\limits_{j=1}^{3}\left(H(i\lambda_j)+\frac{1}{2}\pi |\lambda_j|\right)+\frac{1}{2}\log(\pi\mu b^2)
+ \frac{1}{Q^2}\,{\rm const.}
\right],
\end{eqnarray}
where the constant on the r.h.s. is independent of $z_j$, $\lambda_j$ and $\pi\mu b^2$.
Comparing (\ref{DOZZlimit}) and (\ref{cl}) we see that the classical limit
of the DOZZ structure constant differs from the classical three-point action
by an additional imaginary term. As has been observed in \cite{Hadasz:2003he}
this inconsistency occurs due to the fact that the classical Liouville action
is by construction symmetric with respect to the reflection
$\alpha\to Q-\alpha$, $(\lambda\to -\lambda)$
whereas the DOZZ three-point function is not.
Under this reflection the DOZZ three-point function changes according to the formula
\cite{Zamolodchikov:1995aa}:
$$
C\left(Q-\alpha_1, \alpha_2, \alpha_3\right)\;=\;
{\cal S}(i\alpha_1 - iQ/2)C\left(\alpha_1, \alpha_2, \alpha_3\right),
$$
where
\begin{equation}
\label{reflection}
{\cal S}(x)\;=\;-\left(\pi\mu\gamma(b^2)\right)^{-2ix/b}
\frac{\Gamma(1+2ix/b)\Gamma(1+2ixb)}{\Gamma(1-2ix/b)\Gamma(1-2ixb)}
\end{equation}
is the so-called {\it reflection amplitude} \cite{Zamolodchikov:1995aa}.
The discrepancy between (\ref{DOZZlimit}) and (\ref{cl}) can
be overcome if we consider the \textit{symmetric} three-point function
$\tilde C\left(\alpha_1, \alpha_2, \alpha_3\right)$ \cite{Hadasz:2003he}:
\begin{eqnarray}\label{sDOZZ}
\tilde C\left(\alpha_1, \alpha_2, \alpha_3\right) &\equiv&
\left[\prod\limits_{j=1}^{3}\sqrt{{\cal S}(i\alpha_j - iQ/2)}\right]
C\left(\alpha_1, \alpha_2, \alpha_3\right)
\end{eqnarray}
instead of $C\left(\alpha_1, \alpha_2, \alpha_3\right)$.
Indeed, taking into account the classical limit of the
reflection amplitude for $\lambda\in \mathbb{R}$:
$$
\log {\cal S}\!\left(-\frac{\lambda}{2b}\right)\;\sim\; \frac{2i}{b^2}\lambda
\left(1-\log|\lambda|+\frac{1}{2}\log(\pi\mu b^2)\right)
$$
one can easily verify that the symmetric three-point function in the
limit $b\to 0$ behaves as follows \cite{Hadasz:2003he}
\begin{eqnarray}\label{symmDOZZ}
\tilde C\!\left(\alpha_1, \alpha_2, \alpha_3\right) &\sim &
\exp\left\lbrace -\frac{1}{b^2}
\left[\sum\limits_{\sigma_1, \sigma_2 = \pm}
F\left( \frac{1+i\lambda_1}{2} + \sigma_1 \frac{i\lambda_2}{2} + \sigma_2 \frac{i\lambda_3}{2}\right)\right.\right.
\nonumber\\
&&\left.\left.+\sum\limits_{j=1}^{3}\left(H(i\lambda_j)+\frac{1}{2}\pi |\lambda_j|\right)+\frac{1}{2}\log(\pi\mu b^2)
+{\rm const.}\right]
\right\rbrace\nonumber
\\
&=&
\exp\left\lbrace-\frac{1}{b^2}\,S_{\rm L}^{(3)}(\lambda_1, \lambda_2, \lambda_3)\right\rbrace,
\end{eqnarray}
where $\alpha_j$, $j=1,2,3$ are given by (\ref{alphas}).

Hence, in order to obtain consistent semiclassical
one-point function from the quantum one it is more convenient to take
as a starting point the quantum one-point function
with the \textit{symmetric} DOZZ structure constant:
\begin{eqnarray}
\label{L1pointA2}
\left\langle\,\mathbf{V}_{\!\!\beta}(1)\,\right\rangle_{\tau}^{\rm sym}&\equiv&
{\rm Tr}_{\cal H}\left(q^{L_0 - c/24} {\bar q}^{{\bar L}_0 - c/24}
\,\mathbf{V}_{\!\!\beta}(z,\bar z)\right)\Big|_{z=1}
\\
\label{L1pointB2}
&=&
\int\limits_{\mathbb{R}^{+}}d{\rm P}\, \tilde C\!\left(Q-\alpha_{\rm P},\beta, \alpha_{\rm P}\right)
\left|{\cal F}^{\Delta_\beta}_{c,\Delta_{\alpha_{\rm P}}}(q)\right|^2,
\end{eqnarray}
where $c$, $\Delta_\beta$, $\Delta_{\alpha_{\rm P}}$  are given by (\ref{para2})-(\ref{para3}).
The operator:
\begin{equation}\label{rescaled}
\mathbf{V}_{\!\!\beta}\;=\;\sqrt{{\cal S}(i\beta-iQ/2)}\,{\sf V}_{\beta}
\end{equation}
in (\ref{L1pointA2}) is the primary Liouville vertex operator ${\sf V}_{\beta}$ rescaled by
the square root of the reflection amplitude (\ref{reflection}).
Also the primary vertex operators ${\sf V}_{\!\alpha_{\rm P}}$'s which
generate intermediate highest weight states
are assumed to be rescaled by the square root of the reflection
amplitude,  in accordance with (\ref{rescaled}).

Now, for heavy insertions (\ref{betaaa})-(\ref{alphaa}) using (\ref{symmDOZZ}) one gets
$$
\tilde C\!\left(Q-\alpha_{\rm P}, \beta, \alpha_{\rm P}\right) \;=\;
\tilde C\!\left(\alpha_{\rm P}, \beta, \alpha_{\rm P}\right) \;\;\stackrel{b\to 0}{\sim}\;\;
{\rm e}^{-\frac{1}{b^2} S_{\rm L}^{(3)}(2p,-i\xi,2p)}.
$$

It is reasonable to assume that the torus one-point block has the $b\to 0$
asymptotic similar to the DOZZ three-point function, i.e. given by the eq.~(\ref{classT}).

Finally, in the limit $b\to 0$ the integral (\ref{L1pointB2})
with heavy insertions is determined by the
saddle point approximation, i.e., by the critical value of the function:
\begin{eqnarray}\label{preAction}
\widehat{S}(\xi, p; q) &=& S_{\rm L}^{(3)}(2p, -i\xi, 2p)
-f^{\tilde\delta}_{\delta}(q)-
{\bar f}^{\tilde\delta}_{\delta}(\bar q)\nonumber
\\
&=&
S_{\rm L}^{(3)}(2p, -i\xi, 2p)-2{\rm Re} f^{\tilde\delta}_{\delta}(q),
\end{eqnarray}
where $\delta=\frac{1}{4}+p^2$ and $\tilde\delta=\frac{1}{4}\left(1-\xi^2\right)$.
Indeed, when $b\to 0$  then the symmetric one-point function behaves as follows
$$
\left\langle\,\mathbf{V}_{\!\!\beta}(1)\,\right\rangle_{\tau}^{\rm sym}\;\sim\;
{\rm e}^{-\frac{1}{b^2}\,S^{\rm torus}_{\rm L}(\xi; q)},
$$
where
$
S^{\rm torus}_{\rm L}(\xi; q) = \widehat{S}(\xi, p_{\ast}; q)
$
and the {\it saddle point momentum} $p_{\ast}=p_{\ast}(\xi,q,\bar q)$ is determined by the equation:
\begin{eqnarray}\label{saddlemomentum}
&& \frac{\partial}{\partial p}\, \widehat{S}(\xi, p; q)\Big|_{p=p_{\ast}} = \;0
\;\;\;\;\;\;\;\;\stackrel{(\ref{preAction})}{\Longrightarrow}\nonumber
\\
&& \frac{\partial}{\partial p}\, S_{\rm L}^{(3)}(2p, -i\xi, 2p)\Big|_{p=p_{\ast}} =\;
2{\rm Re}\frac{\partial}{\partial p}\,f^{\tilde\delta}_{\frac{1}{4}+p^{2}}(q)\Big|_{p=p_{\ast}}.
\end{eqnarray}
One thus gets the factorization
\begin{eqnarray}\label{factorization}
S^{\rm torus}_{\rm L}(\xi; q) &=& \widehat{S}(\xi, p_{\ast}; q)\nonumber
\\
&=& S_{\rm L}^{(3)}(2p_{\ast}, -i\xi, 2p_{\ast})
-2{\rm Re} f^{\tilde\delta}_{\frac{1}{4}+p^{2}_{\ast}}(q).
\end{eqnarray}

As a final remark in this paragraph let us note that the modular invariance
of the torus Liouville one-point function \cite{Sonoda,HJSmodular} implies
the {\it classical modular bootstrap} equation:
$$
S^{\rm torus}_{\rm L}\left(\xi; {\rm e}^{2\pi i \tau}\right) \;=\;
S^{\rm torus}_{\rm L}\left(\xi; {\rm e}^{-2\pi i \frac{1}{\tau}}\right)-2\tilde\delta\log|\tau|.
$$

\subsection{Accessory parameter from torus one-point classical action}
\label{nvd}
Consider the null fields \cite{Belavin:1984vu}
\begin{equation}
\label{zerowe2B} \chi_{\pm}(z)\;=\;\left(\widehat{L}_{-2}(z) -
\frac{3}{2(2\Delta_{\pm} +1)}
\,\widehat{L}_{-1}^{\,2}(z)\right){\sf V}_{\pm}(z)
\end{equation}
which correspond to the null vectors
\begin{equation}
\label{rel1} |\,\chi_{\pm}\,\rangle  \;=\;  \left(L_{-2} -
\frac{3}{2(2\Delta_{\pm} +1)}\,L_{-1}^2\right)
|\,\Delta_{\pm}\,\rangle
\end{equation}
appearing on the second level of the Verma module. The operators ${\sf V}_{\pm}$
in (\ref{zerowe2B}) are the primary degenerate
fields with the following conformal weights:
$$
\Delta_{+}  \;=\; -\frac12 -\frac34 b^2 , \hskip 1.5cm
\Delta_{-}  \;=\; -\frac12 -\frac3{4 b^2}.
$$
The correlation functions with null fields must vanish.
In particular, for the two-point function on a torus\footnote{with periods $1$ and $\tau$}
with the null field $\chi_{+}\equiv\chi_{\alpha_+ = -\frac{b}{2}}$ one has
\begin{eqnarray}
\label{torus2P1}
\left\langle\chi_{+}(z)\textsf{V}_{\beta}(w)\right\rangle_\tau &=&
\left\langle \widehat{L}_{-2}(z)\textsf{V}_{+}(z)\textsf{V}_{\beta}(w)\right\rangle_\tau
\nonumber\\
&+&\frac{1}{b^2}\,
\left\langle \widehat{L}_{-1}^{2}(z)\textsf{V}_{+}(z)\textsf{V}_{\beta}(w)\right\rangle_\tau
\;=\; 0.
\end{eqnarray}
Using the torus Ward identities \cite{EO} one can convert the condition (\ref{torus2P1}) to
the second order differential equation:
\begin{eqnarray}\label{torus2point}
&&\left[\frac{1}{b^2}\,\frac{\partial^{2}}{\partial z^2}+
\left(2\Delta_+ \eta_1 + 2\eta_1 z \frac{\partial}{\partial z}\right) + \Delta_\beta
\left(\wp(z-w)+2\eta_1\right)\right.
\\[3pt]
&&\hspace{1cm}\left.+
\left(\zeta(z-w)+2\eta_1 w\right)
\frac{\partial}{\partial w}\right]\left\langle\, {\sf V}_{+}(z){\sf V}_{\beta}(w)\right\rangle_{\!\tau}
\;=\;
-\frac{2\pi i}{Z(\tau)}\;\frac{\partial}{\partial\tau}
\left[Z(\tau)\left\langle\,{\sf V}_{+}(z){\sf V}_{\beta}(w)\right\rangle_{\!\tau} \right],\nonumber
\end{eqnarray}
where $Z(\tau)$ is a partition function and
\begin{eqnarray*}
\zeta(z|\tau)&=&\partial_{z}\log\theta_{1}(z|\tau)+2\eta_1 z,\\
\wp(z)&=&-\partial_{z}\zeta(z|\tau),\\
\eta(\tau)&=&{\rm e}^{2i\pi\tau/24}\prod\limits_{n>0}(1-{\rm e}^{2i\pi\tau n})
\;=\;q^{\frac{1}{24}}\prod\limits_{n>0}(1-q^{n}),\\
\eta_1 &=& (2\pi)^2\left(\frac{1}{24}-\sum\limits_{n=1}^{\infty}\frac{n {\rm e}^{2\pi i n\tau}}
{1-{\rm e}^{2\pi i n\tau}}\right)
\;=\; -2\pi i \partial_{\tau}\log\eta(\tau).
\end{eqnarray*}
Let us introduce
\begin{eqnarray*}
\zeta_{*}(z|\tau)&=&\zeta(z|\tau)-2\eta_1 z \;=\; \partial_{z}\log\theta_{1}(z|\tau),\\
\wp_{*}(z)&=&-\partial_{z}\zeta_{*}(z|\tau)\;=\;\wp(z)+2\eta_1.
\end{eqnarray*}
For $w=0$ from (\ref{torus2point}) one gets
\begin{eqnarray}\label{torus2pointB}
&&\left[-\frac{1}{b^2}\,\frac{\partial^{2}}{\partial z^2}+
\zeta_{*}(z)\,\frac{\partial}{\partial z} - \Delta_\beta\,\wp_{*}(z)-2\Delta_+ \eta_1
\right]\left\langle\, {\sf V}_{+}(z){\sf V}_{\beta}(0)\right\rangle_{\!\tau}
\nonumber\\[3pt]
&&\hspace{1cm}\;=\;
2\pi i\;\left(\frac{\partial}{\partial\tau}\log Z(\tau)\right)
\left\langle\,{\sf V}_{+}(z){\sf V}_{\beta}(0)\right\rangle_{\!\tau}
-4\pi^2 q\frac{\partial}{\partial q}
\left\langle\,{\sf V}_{+}(z){\sf V}_{\beta}(0)\right\rangle_{\!\tau}.
\end{eqnarray}
Recall, that the primary vertex operators in the two-point function
above are defined on the cylinder ${\sf V}_{\sigma}\equiv{\sf V}_{\sigma}^{\rm cyl}$.
These operators are related to the operators living
on the plane (used in the previous paragraph) in the standard way
${\sf V}_{\sigma}^{\rm cyl}(z,\bar z)={\rm e}^{z\Delta_{\sigma}}
{\rm e}^{\bar z\Delta_{\sigma}}{\sf V}_{\sigma}^{\mathbb{C}}({\rm e}^z,{\rm e}^{\bar z})$.

One can think of (\ref{torus2pointB}) as the equation obeyed by the
Liouville two-point function on the torus where the ``alpha'' $\alpha_+$ or equivalently the
conformal weight $\Delta_{\alpha_+}$ of the operator located at $z$
``has been continued'' to the degenerate value.\footnote{The idea
which makes use of the degenerate representation of the Virasoro algebra in the Liouville
field theory is not new. In particular, such trick has been used by Teschner in order
to re-derive the DOZZ formula \cite{T} (for reviews, see \cite{JT1,Nakayama}).}
If so, let us multiply both sides of the eq.~(\ref{torus2pointB}) by
$\left[{\cal S}(i\bar\alpha_{\rm P} - iQ/2)\,
{\cal S}(i\beta - iQ/2)\,
{\cal S}(i\alpha_{\rm P} - iQ/2)\right]^{\frac{1}{2}}=:{\cal R}$.
Next, let us assume that $\alpha_{\rm P}$ and $\beta$ are heavy.
On the other hand
for $b\to 0$ the operator ${\sf V}_{+}$ remains {\it light} $(\Delta_+ = {\cal O}(1))$
and its presence in the correlation function has no influence on the classical dynamics.
Also $\log Z(\tau)$ is of order ${\cal O}(1)$. Hence, one can expect that
for $b\to 0$  \footnote{Eq.~(\ref{approx}) is justified by the well known semiclassical 
behavior of Liouville correlators with heavy and light vertices 
on the sphere, see for instance \cite{Witten}.
It is reasonable to expect that the same holds on the cylinder, cf.~\cite{KashaniPoor:2012wb}.}
\begin{eqnarray}\label{approx}
{\cal R}
\left\langle\,{\sf V}_{+}^{\rm cyl}(z){\sf V}_{\beta}^{\rm cyl}(0)\right\rangle_{\!\tau}
&\sim &
\Psi(z)\, {\cal R} \left\langle\,{\sf V}_{\beta}^{\rm cyl}(0)\right\rangle_{\!\tau}
\;=\;
\Psi(z)\, {\cal R} \left\langle\,{\sf V}_{\beta}^{\mathbb{C}}(1)\right\rangle_{\!\tau}\nonumber
\\
&=&
\Psi(z) \left\langle\,\mathbf{V}_{\!\!\beta}^{\mathbb{C}}(1)\,\right\rangle_{\tau}^{\rm sym}
\;\sim\;
\Psi(z)\,{\rm e}^{-\frac{1}{b^2}\,S^{\rm torus}_{\rm L}(\xi; q)}.
\end{eqnarray}
After substituting (\ref{approx}) into the eq. (\ref{torus2pointB}) and
taking the limit $b\to 0$ one gets
$$
\frac{\partial^{2}}{\partial z^2}\,\Psi(z)+
\left(\tilde\delta\,\wp_{*}(z)+4\pi^2 q\frac{\partial}{\partial q}
\,S^{\rm torus}_{\rm L}(\xi; q)\right)\Psi(z)\;=\;0.
$$
From eqs. (\ref{saddlemomentum}) and (\ref{factorization}) we have
\begin{eqnarray*}
\frac{\partial}{\partial q}\,S^{\rm torus}_{\rm L}(\xi; q)
&=& \frac{\partial}{\partial q}\,\widehat{S}(\xi, p_{\ast}(q); q)
\\
&=&
\frac{\partial}{\partial p}\,\widehat{S}(\xi, p; q)
\Big|_{p=p_{\ast}(q)}\,
\frac{\partial p_{\ast}(q)}{\partial q}
+
\frac{\partial}{\partial q}\,\widehat{S}(\xi, p_{\ast}; q)
\\
&=& -\frac{\partial}{\partial q}\ f^{\tilde\delta}_{\frac{1}{4}+p^{2}_{\ast}}(q).
\end{eqnarray*}
Therefore,
$$
\frac{\partial^{2}}{\partial z^2}\,\Psi(z)-
\left[-\tilde\delta\wp(z)+4\pi^2
q\frac{\partial}{\partial q}\,f^{\tilde\delta}_{\frac{1}{4}+p^{2}_{\ast}}(q)
-2\tilde\delta\eta_1\right]\Psi(z) \;=\;0.
$$
Now, one can identify the parameters $\kappa$ and $\mathrm{B}$ appearing
in the Lam\'{e} equation (\ref{LameW}) as follows
\begin{equation}\label{AB}\boxed{
\;\;\kappa\;=\; -\,\tilde\delta,
\;\;\;\;\;
\mathrm{B}\;=\; 4\pi^2
q\frac{\partial}{\partial q}\,f^{\tilde\delta}_{\frac{1}{4}+p^{2}_{\ast}}(q)
-2\tilde\delta\eta_1\,.}
\end{equation}
Then, for $-\kappa=\tilde\delta=\frac{1}{4}$ (parabolic singularity) the
accessory parameter $\mathrm{C}$ which occurs in the version
(\ref{LameWII}) of the Lam\'{e} equation explicitly reads as follows
\begin{equation}\boxed{
\mathrm{C}\;=\;-4\mathrm{B}\;=\;-\,16\pi^2\,
q\frac{\partial}{\partial q}\,f^{\frac{1}{4}}_{\frac{1}{4}+p^{2}_{\ast}}(q)
+2\eta_1\,.}
\end{equation}
More in general, i.e. for the elliptic singularities $-\kappa=\tilde\delta\in(0, \frac{1}{4})$
one gets
\begin{equation}\boxed{
\mathrm{C}_{\rm ell}\;=\;\frac{\mathrm{B}}{\kappa}\;=\;-\,4\pi^2\,\tilde\delta^{-1}\,
q\frac{\partial}{\partial q}\,f^{\tilde\delta}_{\frac{1}{4}+p^{2}_{\ast}}(q)
+2\eta_1\,.}
\end{equation}
Let us note that $\eta_1$ depends on the modular parameter $\tau$ according
to the formula
$\eta_1 = \frac{4\pi^2}{24}\,E_{2}(\tau)$
where $E_{2}(\tau)$ is the second
Eisenstein series. Hence, finally one can express $\mathrm{B}$
(and $\mathrm{C}$, $\mathrm{C}_{\rm ell}$) in terms of functions depending
on $\tau$:
\begin{equation}\label{AB2}\boxed{
\frac{\mathrm{B}(\tau)}{4\pi^2}\;=\;
q\frac{\partial}{\partial q}\,f^{\tilde\delta}_{\frac{1}{4}+p^{2}_{\ast}}(q)
-\frac{\tilde\delta}{12}\,E_{2}(\tau)\,.}
\end{equation}

\section{Lam\'{e} accessory parameter from ${\cal N}=2^{*}$ gauge theory}
\subsection{Accessory parameter from twisted superpotential}
In order to compute the torus classical block $f^{\tilde\delta}_{\delta}(q)$
entering the expressions for $\mathrm{B}$ and/or $\mathrm{C}$, $\mathrm{C}_{\rm ell}$
one can exploit the ``chiral'' AGT relation on the torus and the correspondence
between the classical limit of the conformal blocks and the Nekrasov-Shatashvili
limit of the Nekrasov instanton partition functions.

The ``chiral'' AGT relation on the torus identifies
the torus quantum block with the Nekrasov instanton
partition function \cite{N,NekraOkun} of the ${\cal N}=2^{\ast}$, ${\rm SU}(2)$ gauge
theory (which equals to
$
[{\cal Z}_{\rm inst}^{{\rm U(1)}}]^{-1}
\times {\cal Z}_{\rm inst}^{{\rm U(2)}}
$
as it is written in the second line of the equation below):
\begin{eqnarray}
\label{AGTtorus}
{q\,}^{\frac{c}{24}-\Delta}
{\cal F}_{c,\Delta}^{\Delta_\beta}(q)&=&
\,
\mathcal{Z}^{{\cal N}=2^{\ast},{\rm SU(2)}}_{{\rm
inst}}(q,a,\mu,\epsilon_1, \epsilon_2)\nonumber
\\
&=&\left(\frac{\eta(q)}{q^{\frac{1}{24}}}\right)^{1-2\Delta_\beta}
\,\mathcal{Z}^{{\cal N}=2^{\ast},{\rm U(2)}}_{{\rm
inst}}(q,a,\mu,\epsilon_1, \epsilon_2).
\end{eqnarray}
In eq.~(\ref{AGTtorus})
$\eta(q)= q^{\frac{1}{24}}\prod_{n=1}^{\infty}(1-q^{\,n})$
is the Dedekind $\eta$-function.
The torus block parameters, namely the external conformal weight
$\Delta_\beta$, the intermediate weight $\Delta$ and the Virasoro
central charge $c$ can be expressed in terms of the ${\cal N}=2^{\ast}$, ${\rm SU}(2)$
super-Yang-Mills theory parameters as follows
\begin{eqnarray}\label{paraTorus1}
c=1+6\frac{(\epsilon_1+\epsilon_2)^2}{\epsilon_1\epsilon_2} \equiv
1+6Q^2
&\;\;\;\Leftrightarrow\;\;\; & b=\sqrt{\frac{\epsilon_2}{\epsilon_1}},
\\\label{paraTorus2}
\Delta_\beta= - \frac{\mu(\epsilon_1+\epsilon_2+\mu)}{\epsilon_1
\epsilon_2} &\;\;\;\Leftrightarrow\;\;\; & \beta=-\frac{\mu}{\sqrt{\epsilon_1 \epsilon_2}},
\\\label{paraTorus3}
\Delta=\frac{(\epsilon_1+\epsilon_2)^2-4a^2}{4\epsilon_1\epsilon_2}
&\;\;\;\Leftrightarrow\;\;\; &
{\rm P}=\frac{ia}{\sqrt{\epsilon_1 \epsilon_2}}.
\end{eqnarray}
Above $\mu$ is the mass of the adjoint
hypermultiplet, $a$ is the vacuum expectation value of the complex scalar of the gauge
multiplet and $\epsilon_1, \epsilon_2$ are $\Omega$-background
parameters. The relation (\ref{AGTtorus}) is understood as an equality between the coefficients
of the expansions of both sides in powers of $q$. For the torus conformal block such
expansion has been introduced in eqs.~(\ref{torusblock})-(\ref{torusCoeff}).
For the definition of the instanton partition function appearing in (\ref{AGTtorus}), see
subsection \ref{SubsectionNSlimit}. The identity (\ref{AGTtorus}) has been proved by Fateev
and Litvinov \cite{FL}. They have shown that the coefficients of the expansions of both
sides of (\ref{AGTtorus}) obey the same recurrence relation.

Let us note that the relation (\ref{AGTtorus}) holds for the heavy conformal weights.
Indeed, from (\ref{paraTorus1})-(\ref{paraTorus3}) we have
\begin{eqnarray}\label{tildedelta}
\tilde\delta &=& \lim\limits_{b\to 0} b^2 \Delta_{\beta}
\;=\;
-\lim\limits_{\epsilon_2\to 0}\frac{\epsilon_2}{\epsilon_1}
\frac{\mu(\epsilon_1+\epsilon_2+\mu)}{\epsilon_1 \epsilon_2}
\;=\;
-\frac{\mu}{\epsilon_1}\left(\frac{\mu}{\epsilon_1}+1\right),
\\\label{delta}
\delta &=&\lim\limits_{b\to 0} b^2 \Delta
\;=\;
\lim\limits_{\epsilon_2\to 0}\frac{\epsilon_2}{\epsilon_1}
\frac{(\epsilon_1+\epsilon_2)^2-4a^2}{4\epsilon_1\epsilon_2}
\;=\;
\frac{1}{4}-\frac{a^2}{\epsilon_{1}^{2}}.
\end{eqnarray}
Hence, one can consider the limit $b\to 0$ of the AGT relation
(\ref{AGTtorus}). The limit $b\to 0$
corresponds to $\epsilon_2\to 0$ ($\epsilon_1={\rm const.}$),
in accordance with (\ref{paraTorus1}). As has been observed by
Nekrasov and Shatashvili \cite{NekraSha} if $\epsilon_2\to 0$  while $\epsilon_1$
is kept finite the Nekrasov instanton partition function
has the following asymptotical behavior:
\begin{equation}\label{NSlimit}
{\cal Z}_{{\rm inst}}(\,\cdot\,,\epsilon_1, \epsilon_2)
\stackrel{\epsilon_2\to 0}{\sim}
\exp\left\lbrace\frac{1}{\epsilon_2}\,
{\cal W}_{\rm inst}(\,\cdot\,,\epsilon_1)\right\rbrace.
\end{equation}
${\cal W}_{\rm inst}(\,\cdot\,,\epsilon_1)$ is the instanton
contribution to the so-called effective twisted superpotential
of the corresponding two-dimensional gauge theory
restricted to the two-dimensional $\Omega$-background.
Twisted superpotentials play also a prominent role in
already mentioned Bethe/gauge correspondence \cite{NekraSha,NekraSha2,NekraSha3}.

Therefore, taking into account the semiclassical
asymptotic (\ref{classT}) of the torus quantum
block and
(\ref{NSlimit}) one can get from (\ref{AGTtorus})
the ``classical version'' of the torus AGT relation:
\begin{eqnarray}\label{classAGT}
f^{\tilde\delta}_{\delta}(q) &=&
\left(\delta - \frac{1}{4} \right)\log q
- 2\tilde\delta\log\left(\frac{\eta(q)}{q^{\frac{1}{24}}}\right)
+
\frac{1}{\epsilon_1}\,
{\cal W}_{\rm inst}^{{\cal N}=2^{*}, {\rm U}(2)}(q,a,\mu,\epsilon_1).
\end{eqnarray}
Using (\ref{classAGT}) and
$
\partial_{\tau}\log\eta(\tau)=\frac{i\pi}{12}\,E_{2}(\tau)
$
one can rewrite the expression (\ref{AB2}) for the parameter $\mathrm{B}$ to the following
form:
\begin{equation}\label{BfromW}\boxed{
\frac{\mathrm{B}(\tau)}{4\pi^2}\;=\;
p_{\ast}^2 + \frac{\tilde\delta}{12}\,
\left(1-2E_{2}(\tau)\right)+\frac{1}{\epsilon_1}\,
q\frac{\partial}{\partial q}\,
{\cal W}_{\rm inst}^{{\cal N}=2^{*}, {\rm U}(2)}(q,a,\mu,\epsilon_1)}
\end{equation}
where
\begin{equation}\label{prameters}
p_{\ast}(\xi,q,\bar q) \;=\; \frac{ia}{\epsilon_1},
\;\;\;\;\;\;\;\;
\tilde\delta \;=\; -\kappa \;=\; -\frac{\mu}{\epsilon_1}\left(\frac{\mu}{\epsilon_1}+1\right)
\;=\;\frac{1}{4}\left(1-\xi^2\right).
\end{equation}
Let us stress that the twisted superpotential
${\cal W}_{\rm inst}^{{\cal N}=2^{*}, {\rm U}(2)}$ is a
computable quantity. The derivation of ${\cal W}_{\rm inst}^{{\cal N}=2^{*}, {\rm U}(2)}$
directly from the Nekrasov instanton partition function
shall be presented in the third subsection. In the next subsection
we confront the calculation  method of the eigenvalue,
which employs the idea of the classical block, with another procedure based on the WKB
analysis, cf. \cite{FL,MirMor,Maruyoshi:2010iu,He:2011zk,KashaniPoor:2012wb}.

\subsection{WKB analysis}
\label{wkb}
Taking into account (\ref{prameters}) one can rewrite the eq.~(\ref{LameW}) to
the following Schr\"{o}dinger-like form:
\begin{equation}\label{SchEq}
-\epsilon_{1}^{2}\,\Psi''(z) + V(z,\epsilon_1)\Psi(z)\;=\;\mathrm{E}\,\Psi(z),
\end{equation}
where
$$
V(z,\epsilon_1)\;=\;\mu\left(\mu+\epsilon_1\right)\wp(z)
=\mu^2 \wp(z)+\epsilon_1\mu\wp(z)
\equiv V_{0}(z)+\epsilon_1 V_{1}(z)
$$
and
\begin{equation}\label{EB}
\mathrm{E}\;=\;-\epsilon_{1}^{2}\,\mathrm{B}.
\end{equation}
Substituting
$$
\Psi(z)\;=\;\exp\left\lbrace -\frac{1}{\epsilon_1}\int^{z}\!\!
P\!\left(x,\epsilon_1\right)dx \right\rbrace
$$
into the eq.~(\ref{SchEq}) one finds
$$
-P^{2}(z,\epsilon_1) + \epsilon_1 P'(z,\epsilon_1) +
 V(z,\epsilon_1)\;=\;\mathrm{E}.
$$
Above equation can be solved iteratively by expansions in $\epsilon_1$:
\begin{eqnarray*}
P(z,\epsilon_1) &=& \sum\limits_{k=0}^{\infty}\epsilon_{1}^{k}\,P_{k}(z),
\;\;\;\;\;\;\;\;\;\;\;\;\;
\;\;\;\;\;\;\;\;
V(z,\epsilon_1) = \sum\limits_{k=0}^{\infty}\epsilon_{1}^{k}\,V_{k}(z).
\end{eqnarray*}
In particular, for lower orders this yields
\begin{eqnarray*}
-P_{0}^{2}+V_{0} &=& \mathrm{E},
\\
-2 P_{0} P_{1} + P_{0}' + V_{1} &=& 0,
\\
-2 P_{0} P_{2} - P_{1}^{2} + P_{1}' + V_2 &=& 0.
\end{eqnarray*}
Note, that in our case $V_{0}(z)=\mu^{2}\wp(z)$,
$V_{1}(z)=\mu\wp(z)$ and $V_{k}(z)=0$ for all $k>1$.

The quasiclassical approximation
$\mathrm{E}\equiv\mathrm{E}|_{{\rm zero}\;{\rm order}\;{\rm in}\;{\rm \epsilon_1}}$
to the energy eigenvalue is determined by the ${\cal A}$-cycle integral
\cite{FL,MirMor,Maruyoshi:2010iu}:
\begin{equation}\label{BSint}
2\pi i a \;=\; \oint\limits_{\cal A} P_{0}dz \;=\;
\oint\limits_{\cal A}\sqrt{V_{0}-\mathrm{E}}\;dz \;=\;
\oint\limits_{\cal A}\sqrt{\mu^2\wp(z)-\mathrm{E}}\;dz.
\end{equation}
Physically, the equation above is nothing but the Bohr-Sommerfeld
quantization condition \cite{MirMor}. It is convenient to introduce
\begin{equation}\label{param}
\mathbb{E}\;=\;\frac{\mathrm{E}}{4\pi^2 a^2},
\;\;\;\;\;\;\;\;\;\;\;\;\;\;\;\;\;\;
\nu\;=\;\frac{\mu^2}{4\pi^2 a^2}
\end{equation}
and rewrite the eq.~(\ref{BSint}) to the form
\begin{equation}\label{BSinst2}
1\;=\;\oint\limits_{\cal A}\sqrt{\mathbb{E}-\nu\wp(z)}\;dz.
\end{equation}
Let us recall that we are working on the torus with periods $1$ and $\tau$
parameterized by the complex coordinate $z\equiv z+1 \equiv z+\tau$.
The ${\cal A}$-cycle here is just the interval $[0,1]$.

Eq.~(\ref{BSinst2}) allows to compute $\mathbb{E}$ as an expansion in $\nu$
with coefficients depending on $q={\rm e}^{2\pi i \tau}$:
\begin{equation}\label{E}
\mathbb{E}\;=\;1 + \mathbb{E}_{1}(q)\,\nu + \mathbb{E}_{2}(q)\,\nu^2 +
\mathbb{E}_{3}(q)\,\nu^3 + \mathbb{E}_{4}(q)\,\nu^4 + \,\ldots\;.
\end{equation}
Indeed, after an expansion of the square root the eq.~(\ref{BSinst2})
becomes
\begin{eqnarray}\label{BSinst3}
1 &=& \oint\limits_{\cal A} dz +
\oint\limits_{\cal A} \frac{1}{2}\left(\mathbb{E}_{1}-\wp(z)\right)\,\nu\,dz
+ \oint\limits_{\cal A} \frac{1}{2}\left(\mathbb{E}_{2}
-\frac{1}{4}\left(\mathbb{E}_1 - \wp(z)\right)^2\right)\,\nu^2\,dz
+ \ldots\;.
\end{eqnarray}
Since
$$
\oint\limits_{\cal A} dz \;=\; 1,
$$
then, the higher terms on the r.h.s. of the eq.~(\ref{BSinst3})
must vanish. Therefore, up to $\nu^4$ one finds \cite{KashaniPoor:2012wb}:
\begin{eqnarray}\label{e1}
\mathbb{E}_{1} &=& \oint\limits_{\cal A} \wp(z) dz = -\frac{\pi^2}{3}E_{2},
\\[10pt]\label{e2}
\mathbb{E}_{2} &=& \oint\limits_{\cal A}\Big[
\frac{1}{4}\wp(z)^2-\frac{1}{2}\mathbb{E}_{1}\wp(z)+\frac{1}{4}\mathbb{E}_{1}^{2} \Big]dz
= \frac{\pi^4}{36}\left(E_{4}-E_{2}^{2}\right),
\\[10pt]
\label{e3}
\mathbb{E}_{3} &=& \oint\limits_{\cal A} \Big[\frac{1}{8}\wp(z)^3
- \frac{3}{8}\mathbb{E}_{1}\wp(z)^2 +
\frac{1}{8}\left(3\mathbb{E}_{1}^{2}-4\mathbb{E}_{2}\right)\wp(z)
+\frac{1}{8}\left(-\mathbb{E}_{1}^{3}+4\mathbb{E}_{1}\mathbb{E}_{2}\right)
\Big] dz\nonumber
\\
&=&
\frac{\pi^6}{540}\left(2E_{6}+3E_{2}E_{4}-5E_{2}^{3}\right),
\\[10pt]\label{e4}
\mathbb{E}_{4} &=& \oint\limits_{\cal A}
\Big[\frac{5}{64}\wp(z)^4 - \frac{5}{16}\mathbb{E}_{1}\wp(z)^3 +
\frac{1}{64}\Big(30\mathbb{E}_{1}^{2}-24\mathbb{E}_{2}\Big)\wp(z)^2\nonumber
\\
&+&
\frac{1}{64}\Big(-20\mathbb{E}_{1}^{3}+48\mathbb{E}_{1}\mathbb{E}_{2}-32\mathbb{E}_{3}\Big)\wp(z)
+\frac{1}{64}\Big(5\mathbb{E}_{1}^{4}-24\mathbb{E}_{1}^{2}\mathbb{E}_{2}+
16\mathbb{E}_{2}^{2}+32\mathbb{E}_{1}\mathbb{E}_{3}\Big)
\Big]dz\nonumber
\\
&=&
\frac{\pi^8}{9072}\left(-35E_{2}^{4}+7E_{2}^{2}E_{4}+10E_{4}^{2}+18E_{2}E_{6}\right),
\end{eqnarray}
where $E_{2n}$, $n=1,2,3$ are the Eisenstein series. In above calculations appear integrals
of the form
$$
\oint\limits_{\cal A} \wp(z)^{n} dz\;=:\; {\k K}_n .
$$
For their explicit computation see appendix B.
Finally, using (\ref{param}), (\ref{E}), (\ref{e1})--(\ref{e4}) and
the $q$-expansions of the Eisenstein series (\ref{E2})--(\ref{E6}) one gets
\begin{eqnarray}\label{EfromWKB}
\frac{1}{4\pi^2}\,\mathrm{E}\Big|_{{\rm zero}\;{\rm order}\;{\rm in}\;{\rm \epsilon_1}}&=&
a^2 - \frac{\mu^2}{12}+\frac{\mu^2\left(4a^2+\mu^2\right)}{2a^2}\;q\nonumber
\\[8pt]
&+&
\frac{\mu^2\left(192 a^6 + 96 a^4 \mu^2 - 48 a^2\mu^4 + 5\mu^6\right)}{32a^6}\;q^2+\ldots\;.
\end{eqnarray}
On the other hand from (\ref{AB2}) and (\ref{EB}) we have
\begin{equation}\label{Efromf}
\frac{1}{4\pi^2}\,\mathrm{E}\Big|_{{\rm zero}\;{\rm order}\;{\rm in}\;{\rm \epsilon_1}}\;=\;
-\lim\limits_{\epsilon_1\to 0}\epsilon_{1}^{2}\frac{\mathrm{B}}{4\pi^2}
\;=\;
-\lim\limits_{\epsilon_1\to 0}\epsilon_{1}^{2}\,\left[
q\frac{\partial}{\partial q}\,f^{\tilde\delta}_{\delta_{*}}(q)
-\frac{\tilde\delta}{12}\,E_{2}(q)\right].
\end{equation}
The torus classical block $f^{\tilde\delta}_{\delta}(q)$ appearing above has the following expansion
\begin{equation}\label{classblock}
f^{\tilde\delta}_{\delta}(q)\;=\;\left(\delta-\frac{1}{4}\right)\log q +
\textbf{f}^{\,\tilde\delta}_{\delta}(q)
\;=\;\left(\delta-\frac{1}{4}\right)\log q + \sum\limits_{n=1}^{\infty}\textbf{f}^{\,\tilde\delta, n}_{\delta}\,q^n
\end{equation}
with coefficients $\textbf{f}^{\,\tilde\delta, n}_{\delta}$ determined by the semiclassical asymptotic
(\ref{classT}) of the quantum block:
$$
\sum\limits_{n=1}^{\infty}\textbf{f}^{\,\tilde\delta, n}_{\delta} q^n \;=\;
\lim\limits_{b\to 0}b^2\log
\left[1+\sum\limits_{n=1}^{\infty}{\cal F}_{1+6Q^2,\Delta}^{\Delta_{\beta},n}\;q^n \right].
$$
For instance,
\begin{equation}\label{f}
\textbf{f}^{\,\tilde\delta, 1}_{\delta} \;=\; \frac{\tilde{\delta }^2}{2 \delta }\,,
\;\;\;\;\;\;\;\;\;\;\;\;\;\;\;\;\;\;\;\;\;\;\;
\textbf{f}^{\,\tilde\delta, 2}_{\delta} \;=\;
\frac{\tilde\delta^2 [24\delta^2 \left(4\delta+1\right)
+ \tilde\delta^2 \left(5\delta-3\right) - 48\tilde\delta\delta^2]}{16 \delta^3 \left(4 \delta +3\right)}\,.
\end{equation}
Using (\ref{tildedelta})--(\ref{delta}), (\ref{E2}) and (\ref{f}) one can check that
(\ref{Efromf}) exactly agrees with (\ref{EfromWKB}).\footnote{Here $\delta_{*}=\frac{1}{4}+p^{2}_{*}$ and
$p^{2}_{*}=-\frac{a^2}{\epsilon_{1}^{2}}$.}

\subsection{Nekrasov--Shatashvili limit}\label{SubsectionNSlimit}
A goal of this subsection is to compute an instanton contribution to the so-called effective twisted
superpotential of the ${\cal N}=2^*$ U(2) gauge theory. We closely follow here the method of the calculation
developed by Poghossian in ref.~\cite{Pogho1}.

Consider the instanton part of the
Nekrasov partition function of the ${\cal N}=2$
supersymmetric ${\rm U}(2)$ gauge theory with an adjoint
hypermultiplet (the ${\cal N}=2^{*}$ theory) \cite{N,NekraSha}:
\begin{eqnarray}\label{inst}
{\cal Z}_{\rm inst}^{{\cal N}=2^{*}, {\rm U}(2)} &=&
1+\sum\limits_{k=1}^{\infty}\frac{q^k}{k!}
\left(\frac{\epsilon_1+\epsilon_2}{\epsilon_1 \epsilon_2}\right)^k\,{\cal Z}_k
\nonumber
\\
&=&
1+\sum\limits_{k=1}^{\infty}\frac{q^k}{k!}
\left(\frac{\epsilon_1+\epsilon_2}{\epsilon_1 \epsilon_2}\right)^k\,
\oint\frac{d\phi_1}{2\pi i}\ldots\oint\frac{d\phi_k}{2\pi i}\;\Omega_k,
\end{eqnarray}
where
\begin{eqnarray*}
\Omega_k &=& \prod\limits_{I=1}^{k}
\frac{P(\phi_I -\mu)P(\phi_I + \mu +\epsilon_1 +\epsilon_2)}
{P(\phi_I)P(\phi_I + \epsilon_1 + \epsilon_2)}
\\
&\times &
\prod_{\begin{array}{c}\scriptstyle I,J=1\\[-7pt]\scriptstyle I \neq J \end{array}}^{k}
\frac{\phi_{IJ}\left(\phi_{IJ}+\epsilon_1+\epsilon_2\right)
\left(\phi_{IJ}+\mu+\epsilon_1\right)
\left(\phi_{IJ}+\mu+\epsilon_2\right)}
{\left(\phi_{IJ}+\epsilon_1\right)\left(\phi_{IJ}+\epsilon_2\right)
\left(\phi_{IJ}+\mu\right)\left(\phi_{IJ}+\mu+\epsilon_1+\epsilon_2\right)},
\end{eqnarray*}
$\phi_{IJ}=\phi_I - \phi_J$ and $P(x)=(x-a_1)(x-a_2)$.
We will assume that $\mu, a_u, \epsilon_1, \epsilon_2\in\mathbb{R}$.
The poles which contribute to the integral (\ref{inst}) are at
\begin{equation}
\label{poles}
\phi_I=\phi_{uij} = a_u + (i-1)\epsilon_1 + (j-1)\epsilon_2,
\;\;\;\;\;\;\;\;\;\;
u=1,2.
\end{equation}
Recall, that these poles are in correspondence with  pairs
${\sf Y} = \left( Y_{1}, Y_{2} \right)$
of Young diagrams
with total number of boxes $|{\sf Y}|=|Y_1|+|Y_2|=k$.
The index $i$ parameterizes the columns whereas $j$
runs over the rows of the diagram $Y_u$.
The parameters $\epsilon_1$, $\epsilon_2$ describe a size
of a box $(i,j)\in Y_u$ in horizontal, vertical direction respectively.
The instanton sum
over $k$ in (\ref{inst}) can be rewritten as a sum over a pairs
of Young diagrams as follows:
$$
{\cal Z}_k = \sum_{{\sf Y}\atop |{\sf Y}|=k} {\cal Z}_{\sf Y}.
$$
The contributions ${\cal Z}_{\sf Y}$ to the instanton sum correspond to
those obtained by performing (in some specific order)
the contour integrals in (\ref{inst}).

Now we want to calculate the Nekrasov-Shatashvili limit \cite{NekraSha}
$\epsilon_2 \to 0$ ($\epsilon_1$ is kept finite)
of the instanton partition function (\ref{inst}).
Based on the arguments developed by Nekrasov and Okounkov in ref.~\cite{NekraOkun}
it  is reasonable to expect that for vanishingly small values
of $\epsilon_2$ the dominant contribution to the instanton
partition function (\ref{inst}) will occur when
$k\sim \frac{1}{\epsilon_2}$.
Unfortunately, we have found no proof of that mechanism in the general case.
Let us note only that this statement becomes evident in the trivial case in which
${\cal Z}_k = 1$ for all $k=1,2,\ldots$. Indeed, for $\epsilon_2 \to 0$
and $x=\frac{q}{\epsilon_2} \in \mathbb{R}_{>0}$ we have then from
eq.~(\ref{inst}):
\begin{eqnarray*}
{\cal Z}_{\rm inst} &=& \sum\limits_{k=0}^{\infty}\frac{1}{k!}
\left(\frac{q}{\epsilon_2}\right)^k
=\sum\limits_{k=0}^{\infty}\frac{x^k}{k!}\,
= \textrm{e}^x = \frac{\textrm{e}^{x\log x}}{\textrm{e}^{x\log x - x}}
\sim\frac{\textrm{e}^{x\log x}}{\textrm{e}^{\log x!}}=\frac{x^{x}}{x!}.
\end{eqnarray*}
This means that the whole sum is dominated by a single term with
$k\sim x\to \infty$.

Hence, in order to compute the limit $\epsilon_2\to 0$ of
${\cal Z}_{\rm inst}^{{\cal N}=2^{*}, {\rm U}(2)}$
let us find first the leading behavior of $\log \left|q^k \Omega_k
\right|$ for large $k$ (i.e. small values of $\epsilon_2$ and finite $\epsilon_1$).
After simple calculations, using
$\log(x\pm\epsilon_2)=\log(x)\pm\frac{\epsilon_2}{x}+O(\epsilon_{2}^{2})$, one gets
$$
\log\left|q^k \Omega_k\right|\;\sim\; \frac{1}{\epsilon_2}
\;{\cal H}_{\rm inst}^{{\cal N}=2^{*}, {\rm U}(2)},
$$
where
\begin{eqnarray}\label{leading}
{\cal H}_{\rm inst}^{{\cal N}=2^{*}, {\rm U}(2)}
&=&  \epsilon_2 k \log|q|+
\epsilon_2 \sum\limits_{I=1}^{k}\log\left|
\frac{P(\phi_I-\mu) P(\phi_I+\mu+\epsilon_1)}{P(\phi_I) P(\phi_I+\epsilon_1)}\right|\nonumber
\\
&+&\;\;\epsilon^{2}_{2}\!\!\!\sum_{\begin{array}{c}\scriptstyle I,J=1\\[-7pt]\scriptstyle I \neq J \end{array}}^{k}
\!\!\left[
\frac{1}{\phi_{IJ}+\epsilon_1}-\frac{1}{\phi_{IJ}}+\frac{1}{\phi_{IJ}+\mu}-
\frac{1}{\phi_{IJ}+\mu+\epsilon_1}\right].
\end{eqnarray}
In eq.~(\ref{leading}) it is implicitly understood that the poles
$\phi_I$ are obtained from eq.~(\ref{poles})  in the limit $\epsilon_2\to 0$.
Note that in the limit $\epsilon_2\to 0$ the poles
form a continuous distribution (cf. \cite{Fucito}):
\begin{equation}\label{bieguny}
\phi_I = \phi_{u,i}\in\left[x_{ui}^{0}, x_{ui}\right]
\end{equation}
where
\begin{eqnarray*}
x_{ui}^{0} &=& a_u + (i-1)\epsilon_1,
\;\;\;\;\;\;\;\;\;\;
u=1,2,
\;\;\;\;\;\;\;\;\;\;
i=1,\ldots,\infty,
\\
x_{ui} &=& a_u + (i-1)\epsilon_1 +\omega_{ui}\;.
\end{eqnarray*}
In terms of Young diagrams the situation can be explained heuristically as follows.
When $\epsilon_2$ is very small then the number of boxes $k_{ui}$ in the vertical
direction (the number of rows) is very large, however this number
multiplied by $\epsilon_2$, i.e.: $\epsilon_2 k_{ui} = \omega_{ui}$ is
expected to be finite. In other words we obtain a continuous distribution of
rows in the limit under consideration. Then, in order to evaluate (\ref{leading})
at the values (\ref{bieguny}) one can assume that
the summations ``over instantons'' in (\ref{leading}) become continuous in the row index:
\begin{equation}\label{sumI}
\epsilon_2\sum\limits_{I}\To\sum\limits_{u,i}
\int\limits_{x_{ui}^{0}}^{x_{ui}}d\phi_{ui}.
\end{equation}
The limits of integration $x_{ui}^{0}$ and $x_{ui}$ are
the bottom and the top ends of the $i$-th column in $Y_u$ respectively.
Applying eq. (\ref{sumI}) to eq.~(\ref{leading}) one gets
\begin{eqnarray}\label{2}
{\cal H}_{\rm inst}^{{\cal N}=2^{*}, {\rm U}(2)}\left(x_{ui}\right) &=&
\sum\limits_{u,v=1}^{2} \sum\limits_{i,j=1}^{\infty}
\Big[
-F\left(x_{ui}-x_{vj}+\epsilon_1\right) + F\left(x_{ui}-x_{vj}^{0}+\epsilon_1\right)
\nonumber\\
&&\hspace{-110pt}+\;
F\left(x_{ui}^{0}-x_{vj}+\epsilon_1\right) - F\left(x_{ui}^{0}-x_{vj}^{0}+\epsilon_1\right)
+
F(x_{ui}-x_{vj}) - F(x_{ui}-x_{vj}^{0})
\nonumber\\
&&\hspace{-110pt}-\; F(x_{ui}^{0}-x_{vj})+F(x_{ui}^{0}-x_{vj}^{0})
-F(x_{ui}-x_{vj}+\mu)+F(x_{ui}-x_{vj}^{0}+\mu)
\nonumber\\
&&\hspace{-110pt}
+F(x_{ui}^{0}-x_{vj}+\mu)-F(x_{ui}^{0}-x_{vj}^{0}+\mu)
+F(x_{ui}-x_{vj}+\mu+\epsilon_1)
\\
&&\hspace{-110pt}
-F(x_{ui}-x_{vj}^{0}+\mu+\epsilon_1)
-F(x_{ui}^{0}-x_{vj}+\mu+\epsilon_1)+F(x_{ui}^{0}-x_{vj}^{0}+\mu+\epsilon_1)
\Big]
\nonumber\\
%
&&\hspace{-110pt}+ \sum\limits_{u,v=1}^{2} \sum\limits_{i=1}^{\infty}
\Big[F\left(x_{ui}-\mu-a_v\right) - F\left(x_{ui}^{0}-\mu-a_v\right)
+F\left(x_{ui}+\mu-a_v +\epsilon_1\right)
\nonumber\\
&&\hspace{-110pt}
-F\left(x_{ui}^{0}+\mu-a_v +\epsilon_1\right)
-F\left(x_{ui}-a_v\right)+F\left(x_{ui}^{0}-a_v\right)
\nonumber\\
&&\hspace{-110pt}
-F\left(x_{ui}-a_v+\epsilon_1\right)+F\left(x_{ui}^{0}-a_v+\epsilon_1\right)
\Big]
+\sum\limits_{u=1}^{2}\sum\limits_{i=1}^{\infty}
\left(x_{ui}-(i-1)\epsilon_1 - a_u\right)\log |q|\;,\nonumber
\end{eqnarray}
where $F(x)=x(\log |x| - 1)$.

Finally, the Nekrasov instanton
partition function in the limit $\epsilon_2\to 0$ can be represented as
follows:
\begin{equation}\label{path}
{\cal Z}_{\rm inst}^{{\cal N}=2^{*}, {\rm U}(2)} \sim\int \left[\prod\limits_{u,i}dx_{ui}\right]
\,\exp\left\lbrace\frac{1}{\epsilon_2}\,
{\cal H}_{\rm inst}^{{\cal N}=2^{*}, {\rm U}(2)}(x_{ui})\right\rbrace,
\end{equation}
where the ``integral'' is over the infinite set of variables
$\left\lbrace x_{ui} : u=1,2; i=1,\ldots,\infty\right\rbrace$.
As a consequence, the Nekrasov-Shatashvili limit
of ${\cal Z}_{\rm inst}^{{\cal N}=2^{*}, {\rm U}(2)}$ is nothing
but the critical value of ${\cal H}_{\rm inst}^{{\cal N}=2^{*}, {\rm U}(2)}$:
\begin{equation}\label{critvalue}
{\cal W}_{\rm inst}^{{\cal N}=2^{*}, {\rm U}(2)}\;\equiv \;
\lim\limits_{\epsilon_2 \to 0}\epsilon_2 \log {\cal Z}_{\rm inst}^{{\cal N}=2^{*}, {\rm U}(2)}
\;=\;{\cal H}_{\rm inst}^{{\cal N}=2^{*}, {\rm U}(2)}(\hat x_{ui}),
\end{equation}
where $\hat x_{ui}$ denotes the ``critical configuration''  extremizing
the ``free energy'' (\ref{2}).

\subsection{Saddle point equation}
The extremality condition for the ``action''
${\cal H}_{\rm inst}^{{\cal N}=2^{*}, {\rm U}(2)}$ given by (\ref{2}) reads as follows:
\begin{eqnarray*}
&&\Big|q
\Big(\prod\limits_{v,j}
\frac{(x_{ui}-x_{vj}-\epsilon_1)(x_{ui}-x_{vj}^{0}+\epsilon_1)}
{(x_{ui}-x_{vj}^{0}-\epsilon_1)(x_{ui}-x_{vj}+\epsilon_1)}
\frac{(x_{ui}-x_{vj}-\mu)(x_{ui}-x_{vj}^{0}+\mu)}
{(x_{ui}-x_{vj}^{0}-\mu)(x_{ui}-x_{vj}+\mu)}
\\
&\times &\!\!\!\!
\frac{(x_{ui}-x_{vj}+\mu+\epsilon_1)(x_{ui}-x_{vj}^{0}-\mu-\epsilon_1)}
{(x_{ui}-x_{vj}^{0}+\mu+\epsilon_1)(x_{ui}-x_{vj}-\mu-\epsilon_1)}
\Big)
\Big(\prod\limits_{v}
\frac{(x_{ui}-\mu-a_{v})(x_{ui}+\mu+\epsilon_1-a_{v})}
{(x_{ui}-a_{v})(x_{ui}+\epsilon_1-a_v)}
\Big)\Big|=1,
\end{eqnarray*}
where $u,v=1,2$ and $i,j=1,\ldots,\infty$.
This implies that either the following eq.:
\begin{eqnarray}\label{saddle2}
&&-q
\Big(\prod\limits_{v,j}
\frac{(x_{ui}-x_{vj}-\epsilon_1)(x_{ui}-x_{vj}^{0}+\epsilon_1)}
{(x_{ui}-x_{vj}^{0}-\epsilon_1)(x_{ui}-x_{vj}+\epsilon_1)}
\frac{(x_{ui}-x_{vj}-\mu)(x_{ui}-x_{vj}^{0}+\mu)}
{(x_{ui}-x_{vj}^{0}-\mu)(x_{ui}-x_{vj}+\mu)}
\\
&\times &\!\!\!\!
\frac{(x_{ui}-x_{vj}+\mu+\epsilon_1)(x_{ui}-x_{vj}^{0}-\mu-\epsilon_1)}
{(x_{ui}-x_{vj}^{0}+\mu+\epsilon_1)(x_{ui}-x_{vj}-\mu-\epsilon_1)}
\Big)
\Big(\prod\limits_{v}
\frac{(x_{ui}-\mu-a_{v})(x_{ui}+\mu+\epsilon_1-a_{v})}
{(x_{ui}-a_{v})(x_{ui}+\epsilon_1-a_v)}
\Big)=1\nonumber
\end{eqnarray}
or its analog in which $-q$ is replaced by $+q$ are holding.\footnote{In eq.~(\ref{saddle2}) 
and below $q\equiv |q|$.}
Eq.~(\ref{saddle2}) can be regularized assuming that there is an integer
$L$ such that the length of the column $\omega_{ui}$ is equal to
zero for $i>L$.
Analyzing eq.~(\ref{saddle2}) in such a case, i.e. when $j=1,\ldots, L$,
one can observe that the column lengths extremizing the ``free energy''
are of the order $\omega_{ui}\sim {\cal O}(q^i)$.\footnote{
Note that the saddle point equation (or in fact a system of equations) must be solved subject to the condition
that the solution is consistent with the $q$-expansion of the twisted superpotential obtained from the
instanton partition function. This implies that the order of $q$ in the expansions of rescaled column lengths
must correlate with the index $i$. Indeed, first, we have
$
{\cal W}_{\rm inst}\equiv
\lim_{\epsilon_2 \to 0}\epsilon_2 \log {\cal Z}_{\rm inst}
=\sum_{i=1}{\cal W}_i \;q^i.
$
Then,
\begin{equation}\label{WI}
q\frac{d}{dq}\,{\cal W}_{\rm inst} = {\cal W}_1\,q + 2{\cal W}_2\,q^2 +\dots\;
=\widetilde{{\cal W}}_1\,q + \widetilde{{\cal W}}_2\,q^2 +\dots\;.
\end{equation}
On the other hand
\begin{equation}\label{WII}
q\frac{d}{dq}\,{\cal W}_{\rm inst} = \sum\limits_{i=1}\sum\limits_{u}\omega_{ui}=
\sum\limits_{u}\omega_{u1}+\sum\limits_{u}\omega_{u2}+\dots\;,
\end{equation}
where $\omega_{ui}$'s are the column lengths of the critical diagrams (see the calculation
at the beginning of the next subsection). Hence, in order to get from (\ref{WII})
the $q$-expansion consistent with (\ref{WI}) the expansions of $\omega_{ui}$'s must start
from $q^i$.}
For example, at order $q^L$ one can write
\begin{equation}\label{solution}
\hat x_{ui}\equiv x_{ui} = a_u + (i-1)\epsilon_1 +\omega_{ui}(\,q)
=a_u + (i-1)\epsilon_1 +\sum\limits_{n=i}^{L}\omega_{uin}\;q^n.
\end{equation}
Here the symbols $\omega_{uin}$ denote the contributions to the
coefficients $\omega_{ui}$ at the $n$-th order in $q$.
Now it is possible to
solve equation (\ref{saddle2}) starting from $L=1$ and deriving
recursively the  $\omega_{uin}$'s step by step up to desired order.
For instance,
\begin{eqnarray*}
\omega_{111} &=& \frac{\mu(2 a-\mu ) \left(\epsilon_1+\mu \right) \left(2 a+\epsilon_1+\mu \right)}
{2 a \epsilon_1 \left(2 a+\epsilon_1\right)},
\\
\omega_{211} &=& -\frac{\mu(2 a+\mu ) \left(\epsilon_1+\mu \right) \left(-2 a+\epsilon_1+\mu \right)}
{2 a \epsilon_1 \left(2 a-\epsilon_1\right)}.
\end{eqnarray*}

In order to investigate the solution of the saddle point equation in the case when $L\to \infty$
it is helpful to convert it to other equivalent form. Indeed, as has been observed in
\cite{Fucito} (see also \cite{Pogho1}) the eq.~(\ref{saddle2})
can be rewritten in terms of certain ``Y-system'':
\begin{equation}\label{Ysystem}
-q\frac{Y(x_{ui}-\epsilon_1)Y(x_{ui}-\mu)Y(x_{ui}+\mu+\epsilon_1)}
{Y(x_{ui}+\epsilon_1)Y(x_{ui}+\mu)Y(x_{ui}-\mu-\epsilon_1)}
\;=\;1,
\end{equation}
where
\begin{equation}\label{Y}
Y(z) \;=\; \prod\limits_{u=1}^{2}\exp\left\lbrace\frac{z}{\epsilon_1}\,
\psi\left(\frac{a_u}{\epsilon_1}\right)\right\rbrace\prod_{i=1}^{\infty}
\left( 1-\frac{z}{x_{ui}} \right)\exp\left\lbrace \frac{z}{x_{ui}^{0}}\right\rbrace
\end{equation}
and $\psi(z)=\partial_{z}\log\Gamma(z)$.
The product in (\ref{Y}) is convergent for arbitrary $z\in\mathbb{C}$
provided that the column lengths tend to zero for large enough $i$, which is equivalent to
the assumption that $x_{ui}\to x_{ui}^{0}$.
If $\omega_{ui} = 0$ for all $i$, i.e. all column lengths
are zero, then $Y(z)$ becomes
\begin{equation}\label{Y0}
Y_{0}(z) \;=\; \prod\limits_{u=1}^{2}\exp\left\lbrace\frac{z}{\epsilon_1}\,
\psi\left(\frac{a_u}{\epsilon_1}\right)\right\rbrace\prod_{i=1}^{\infty}
\left( 1-\frac{z}{x_{ui}^{0}} \right)\exp\left\lbrace \frac{z}{x_{ui}^{0}} \right\rbrace.
\end{equation}
The functions $Y(z)$, $Y_{0}(z)$ have zeros located at $x_{ui}$ and $x_{ui}^{0}$
respectively.

\subsection{Twisted superpotential, classical block, accessory parameter}
Now we are ready to compute the critical value
of the ``free energy'' (\ref{critvalue}), i.e. the so-called twisted superpotential.
It is convenient first to calculate the derivative of
${\cal W}_{\rm inst}^{{\cal N}=2^{*}, {\rm U}(2)}(\,q, a, m_i; \epsilon_1)$ with
respect to $q$:
\begin{eqnarray}\label{derW}
\frac{\partial}{\partial q}\,{\cal W}_{\rm inst}^{{\cal N}=2^{*}, {\rm U}(2)}
(\,q, a, \mu; \epsilon_1)
&=&
\left(\frac{\partial {\cal H}_{\rm inst}}{\partial x_{ui}}\frac{\partial x_{ui}}{\partial q}
+
\frac{\partial {\cal H}_{\rm inst}}{\partial q}\right)\Big|_{x_{ui}=\hat x_{ui}}
=
\frac{1}{q}\sum\limits_{u,i}\omega_{ui}.
\end{eqnarray}
Above we have used the fact that $\partial {\cal
  H}_{\rm inst}/\partial x_{ui}|_{x_{ui}=\hat x_{ui}} =0$.
Hence, it is easy to realize that the last term
 in (\ref{derW}) coincides with the sum over the column lengths of the
``critical'' Young diagrams. More explicitly, the eq.~(\ref{derW}) reads as follows
\begin{eqnarray}
q\frac{\partial}{\partial q}\,{\cal W}_{\rm inst}^{{\cal N}=2^{*}, {\rm U}(2)}
&=&
\sum\limits_{i}\left(\omega_{1i}(\,q)+\omega_{2i}(\, q)\right)
=
\sum\limits_{i}\left[\sum\limits_{n=i}\left(\omega_{1in} + \omega_{2in}\right)q^n\right]\nonumber
\\[8pt]
&=&
\left[\left(\omega_{111} + \omega_{211}\right) q
+\left(\omega_{112} + \omega_{212}\right) q^2 +\ldots \right]\nonumber
\\[8pt]
&+&
\left[\left(\omega_{122} + \omega_{222}\right) q^2
+\left(\omega_{123} + \omega_{223}\right) q^3 +\ldots\right]
+
\ldots\;.
\end{eqnarray}
Then,
\begin{eqnarray}
\label{Wexp}
{\cal W}_{\rm inst}^{{\cal N}=2^{*}, {\rm U}(2)}
&=&
\left({\omega}_{111}+{\omega}_{211}\right)q \;+\;
\left({\omega}_{112}+{\omega}_{212}+{\omega}_{122}+{\omega}_{222}\right)\frac{q^2}{2}
\;+\;\ldots\;.
\end{eqnarray}

Knowing the coefficients of the extremal lengths of the columns from a solution
of the system of equations (\ref{saddle2}) one gets
\begin{eqnarray}\label{classAGTexp}
\frac{1}{\epsilon_1}\,{\cal W}_{\rm inst}^{{\cal N}=2^{*}, {\rm U}(2)}
&=&
\frac{1}{\epsilon_1}\,\left({\omega}_{111}+{\omega}_{211}\right)q \;+\;\ldots
\;=\;
\frac{1}{\epsilon_1}\,\left(\textbf{f}_{\delta}^{\,\tilde\delta,1}-2\tilde\delta\right)\,q + \ldots\;,
\end{eqnarray}
where $\delta$, $\tilde\delta$ are given by (\ref{tildedelta})--(\ref{delta})
and $\textbf{f}_{\delta}^{\,\tilde\delta,n}$, $n=1,\ldots$ denote coefficients of the torus classical block
(see (\ref{classblock})-(\ref{f})).
Concluding, the eq.~(\ref{classAGTexp}) is nothing but the expansion of both sides of the
``classical'' AGT relation\footnote{Of course,  eqs.~(\ref{classAGT2}) and (\ref{classAGT})
are exactly the same.}:
\begin{equation}\label{classAGT2}
\textbf{f}^{\,\tilde\delta}_{\delta}(q) \;=\;
- 2\tilde\delta\log\left(\frac{\eta(q)}{q^{\frac{1}{24}}}\right)
\;+\;
\frac{1}{\epsilon_1}\,
{\cal W}_{\rm inst}^{{\cal N}=2^{*}, {\rm U}(2)}(q,a,\mu;\epsilon_1).
\end{equation}

As a final conclusion of this subsection let us write down the main result
of the present work. Knowing the classical torus one-point block from
(\ref{classAGT}) and applying eqs.~(\ref{BfromW}) and (\ref{derW}), one arrives at the following
expression of the Lam\'{e} accessory parameter:
\begin{equation}\label{BfromW2}\boxed{
\frac{\mathrm{B}(\tau)}{4\pi^2}\;=\;
p_{\ast}^2(\xi,q,\bar q) + \frac{\tilde\delta}{12}\,
\left(1-2E_{2}(\tau)\right)+\frac{1}{\epsilon_1}\,
\sum\limits_{u,i}\omega_{ui}(q,a,\mu,\epsilon_1)}
\end{equation}
where
$$
p_{\ast}(\xi,q,\bar q) \;=\; \frac{ia}{\epsilon_1},
\;\;\;\;\;\;\;\;
\tilde\delta \;=\; -\kappa \;=\; -\frac{\mu}{\epsilon_1}\left(\frac{\mu}{\epsilon_1}+1\right)
\;=\;\frac{1}{4}\left(1-\xi^2\right).
$$
Hence, we have found that the accessory parameter
$\mathrm{B}$ is related to the sum of column lengths of the ``critical''
Young diagrams. The latter can be rewritten using the contour
integral representation. Indeed, let $\gamma$ denotes
the contour which encloses all the points $\hat x_{ui}$, $x_{ui}^{0}$, $u=1,2$, $r=1,\ldots,\infty$.
Then, as has been noticed by Poghossian in ref.~\cite{Pogho1}, the sum $\sum_{u,i}\omega_{ui}$
can be expressed as follows
$$
\sum_{u,i}\limits\omega_{ui} \;=\;  \sum_{u,i}\left(\hat x_{ui}-x_{ui}^{0}\right)
\;=\;
\oint\limits_{\gamma}
\frac{dz}{2\pi i}\,z\frac{\partial}{\partial z}\log\frac{Y(z)}{Y_{0}(z)}.
$$

\section{Concluding remarks and open problems}
The main result of the present work is the expression of the Lam\'{e} accessory
parameter $\mathrm{B}$ (or equivalently  $\mathrm{A}$, $\mathrm{C}$, $\mathrm{C}_{\rm ell}$)
in terms of the solution of the TBA-like eq.~(\ref{saddle2}).
This equation has been solved by a power expansion in the parameter $q$.
It has been noticed that obtained expression for
$\mathrm{B}$ can be rewritten in terms of a contour integral in which
the integrand is built out of the functions $Y$ and $Y_0$ introduced in eqs.~(\ref{Y})--(\ref{Y0}).
As has been observed in \cite{Pogho1,Fucito} also the TBA-like eq.~(\ref{saddle2}) can
be rewritten in terms of the function $Y$.
Hence, one can relate the problem of finding the Lam\'{e} accessory parameter
to the problem of searching a solution of the functional equation (\ref{Ysystem}).

Another result presented in this paper is a check that the classical limit of
the torus quantum conformal block exists and yields a consistent definition of the
torus classical block. Moreover, it has been verified that the torus classical block
corresponds to the twisted superpotential of the ${\cal N}=2^*$ ${\rm U}(2)$ gauge theory.

Finally, in this paper has been proposed an expression for the
Liouville classical action $S^{\rm torus}_{\rm L}$
on the one-punctured torus. It has been
conjectured that $S^{\rm torus}_{\rm L}$ can be calculated in
terms of the torus classical
conformal block, classical three-point Liouville action and the saddle
point classical intermediate weight (the saddle point momentum).
We leave as an open question (to which we return very soon)
a comparison of this result with the results obtained by Menotti in
\cite{Menotti1,Menotti2,Menotti3,Menotti:2013bka}.

Work is in progress in order to verify formulae (\ref{AB})--(\ref{AB2}) and (\ref{BfromW2}).
First of all let us note  that these formulae pave the way for
numerical studies of the function $\mathrm{B}(\tau)$ and the results of such investigation
can be compared with that obtained by Keen, Rauch and Vasquez in \cite{KRV}.
It is well known that in a certain limit the Lam\'{e}
equation becomes the Mathieu equation. Hence, other possible check beyond the WKB analysis
performed in subsection \ref{wkb} is to verify whether our candidate for
the Lam\'{e} eigenvalue correctly reduces to the Mathieu eigenvalue, cf. \cite{He:2011zk}.\footnote{
Let us stress that we have found an agreement between $\mathrm{B}(\tau)$ and
the energy eigenvalue $\mathrm{E}$ computed by means of the WKB method so far only
for $\epsilon_1=0$. It remains still to calculate the
``quantum corrections'' to $\mathrm{E}$ in the higher powers of
$\epsilon_1$  and compare the result with $\mathrm{B}(\tau)$ obtained from the
classical block.}
As a final remark concerning possible combinatorial tests of our
main result let us note that it would be valuable to check whether one can recover
from eqs.~(\ref{AB2}), (\ref{BfromW}) the expansions of the Lam\'{e}
eigenvalue worked out by M\"{u}ller-Kirsten \cite{MK} (see also \cite{Dunne:1999zc})
and Longmann \cite{Langmann:2004sj,Langmann}.\footnote{Note, that
since the modular properties of the Lam\'{e} accessory parameter are known \cite{KRV},
then the relation (\ref{BfromW}), if it is true, encodes
some information about the
modular transformation properties of the ${\cal N}=2^*$ U(2) twisted superpotential.
This observation fits in an interesting line of research
which aims to answer the question how the S-duality is realized
in the $\Omega$-deformed ${\cal N}=2$ gauge theories, see for instance \cite{Billo1,Billo2}.}

It is a well known fact \cite{KRV} that the Lam\'{e} accessory parameter is
a modular form of weight $2$. In order to answer the question
whether proposed candidate for the Lam\'{e} accessory parameter has
correct modular transformation properties one has to put a heuristic
derivation of eqs.~(\ref{AB2}), (\ref{BfromW}), (\ref{BfromW2}) on a more rigorous
mathematical level. First, it has to be proved that in the classical limit the quantum
conformal blocks behave exponentially.\footnote{It seems to be possible to prove the
classical asymptotics of conformal blocks representing them as the 
Coulomb gas/Dotsenko-Fateev/$\beta$-ensemble
integrals \cite{DS,MMS2,MMS3,MMM2} and applying matrix models technics
\cite{BMT,Dijkgraaf,Sulkowski,Bourgine1,Bourgine2,Bourgine3,MMM,MMS,Morozov,FPplb,FPcjp}.}
Let us stress that there is still no rigorous proof
of convergency of the expansions defining generic quantum conformal blocks.
Secondly, it seems to be possible to derive
the ``classical'' torus AGT relation (\ref{classAGT}) more rigorously
exploiting the Teichm\"{u}ller theory approach to Liouville theory.
Indeed, relations such as (\ref{classAGT}) implicitly appear as a byproduct in 
the proof of the AGT correspondence recently proposed
by Teschner and Vartanov \cite{Vartanov:2013ima}.

\appendix
\section{Special functions}
\renewcommand{\theequation}{A.\arabic{equation}}
\setcounter{equation}{0}
\label{ellipticf}

\subsection*{Jacobi and Weierstrass elliptic functions}
Let $\omega_1$ and $\omega_2$ be two complex numbers whose
ratio is not real. A function which satisfies
$
f(z)=f(z+2\omega_1)=f(z+2\omega_2)
$
for all $z\in D(f(z))$ is called a {\it doubly periodic} function of $z$ with periods
$2\omega_1$ and $2\omega_2$. A doubly periodic function that is
analytic except at its poles and which has no singularities other
than these poles in a finite part of the complex plane is called
an {\it elliptic function}.

Let ${\rm Im}\left(\frac{\omega_2}{\omega_1}\right)\in\mathbb{R}$,
then the points
$0$, $2\omega_1$, $2\omega_1 + 2\omega_2$, $2\omega_2$ when taken in order are the
vertices of a parallelogram, known as the {\it fundamental parallelogram}.
The behaviour of an elliptic function is completely determined by its values
in fundamental parallelogram.

If we consider the points of the period lattice defined as ${\cal L}=\left\lbrace
2m\omega_1 + 2n\omega_2\right\rbrace$, then the four points
$2m\omega_1 + 2n\omega_2$,
$2(m+1)\omega_1 + 2n\omega_2$,
$2(m+1)\omega_1 + 2(n+1)\omega_2$,
$2m\omega_1 + 2(n+1)\omega_2$
are vertices of a similar parallelogram, obtained from the
fundamental parallelogram by a translation without rotations. This parallelogram is called
a {\it period parallelogram}. The complex plane is covered by a system of non-overlapping
period parallelograms.

If we wish to count the number of poles or zeros of an elliptic function in a given period
parallelogram and it happens that certain poles or zeros lie on the boundaries of this parallelogram, then
one can translate the period parallelogram without rotation until no pole or zero
lies on its boundary. Such obtained parallelogram is called a {\it cell}.

In \cite{WW} one can find a list of general properties of elliptic functions. In particular,
one can prove, that (i) the number of poles of an elliptic function in any cell is finite;
(ii) the sum of the residues of an elliptic function at its poles in any cell is zero.

The number of poles of an elliptic function in any cell, counted with multiplicity,
is called the {\it order} of the function. The statements (i)-(ii) written down above imply,
that the order of an elliptic function is necessarily at least equal to $2$. Indeed,
an elliptic function of order $1$ would have a single irreducible pole. If this were
actually a pole its residue would not be zero. Hence, in terms of singularities, the
simplest elliptic functions are those of order 2. These can be divided into two classes:
(I) those which have a single irreducible double pole in each cell at which the residue
is zero; (II) those which have two simple poles in each cell at which the two residues are
equal in absolute value, but of opposite sign.

The {\it Jacobi elliptic functions} are examples of the second class
of elliptic functions of order $2$.
The Jacobi function ${\rm sn}\, u$ is defined by means of the integral
$$
u\;=\;\int\limits_{0}^{x}\frac{dt}{\sqrt{(1-t^2)(1-k^2 t^2)}},
$$
where $k$ is a constant. By inversion of the integral we have $x={\rm sn}\, u$.
From definition follows that ${\rm sn}\, 0 = 0$.

The functions ${\rm cn}\, u$ and ${\rm dn}\, u$ are defined by the identities:
\begin{eqnarray*}
{\rm sn}^2 u + {\rm cn}^2 u \;=\;1,
\;\;\;\;\;\;\;\;\;\;\;\;\;\;
k^2 {\rm sn}^2 u + {\rm dn}^2 u \;=\; 1.
\end{eqnarray*}
It follows that ${\rm cn}\,0={\rm dn}\, 0=1$.

Each of the Jacobi elliptic functions depends on a parameter $k$, called the {\it modulus}.
In order to emphasize this dependence one can write the three functions as ${\rm sn}(u,k)$,
${\rm cn}(u,k)$, ${\rm dn}(u,k)$. An alternative notation:  ${\rm sn}(u|m)$,
${\rm cn}(u|m)$, ${\rm dn}(u|m)$ uses a parameter $m=k^2$.

In accordance with the definition of an elliptic function the Jacobi elliptic functions are
doubly periodic:
\begin{eqnarray*}
{\rm sn}\, u &=& {\rm sn}\!\left(u+4{\cal K}\right)
\;=\;{\rm sn}\!\left(u+4{\cal K}+4i{\cal K}'\right)\;=\;{\rm sn}\!\left(u+2i{\cal K}'\right),
\\
{\rm cn}\, u &=& {\rm cn}\!\left(u+4{\cal K}\right)
\;=\;{\rm cn}\!\left(u+2{\cal K}+2i{\cal K}'\right)\;=\;{\rm cn}\!\left(u+4i{\cal K}'\right),
\\
{\rm dn}\, u &=& {\rm dn}\!\left(u+2{\cal K}\right)
\;=\;{\rm dn}\!\left(u+4{\cal K}+4i{\cal K}'\right)\;=\;{\rm dn}\!\left(u+4i{\cal K}'\right).
\end{eqnarray*}
Periods are expressed in terms of the constants ${\cal K}$ and ${\cal K}'$ defined as follows
$$
{\cal K}\;\equiv\;K(k)\;=\;\int\limits_{0}^{1}\frac{dt}{\sqrt{(1-t^2)(1-k^2 t^2)}},
\;\;\;\;\;\;\;\;\;\;\;\;\;
{\cal K}' \;\equiv\; K(k'),
$$
where $k'$ is the so-called {\it complementary modulus}
defined by the relation $k^2 + k'^2=1$. The integral $K(k)$ is the complete
elliptic integral of the first kind.

The derivatives of the Jacobi elliptic functions are
$$
\frac{d}{d u}\, {\rm sn}\, u = {\rm cn}\, u \,{\rm dn}\, u,
\;\;\;\;\;\;\;\;
\frac{d}{d u}\, {\rm cn}\, u = -\, {\rm sn}\, u \,{\rm dn}\, u,
\;\;\;\;\;\;\;\;
\frac{d}{d u} \,{\rm dn}\, u = -k^2\, {\rm sn}\, u \,{\rm cn}\, u.
$$

For practical reasons it is convenient to introduce a shortened
notation to express reciprocals and quotients of the Jacobi elliptic functions.
The reciprocals are denoted by reversing the orders of the letters of the function:
$$
{\rm ns}\, u \;=\; \frac{1}{{\rm sn}\, u},
\;\;\;\;\;\;\;\;
{\rm nc}\, u \;=\; \frac{1}{{\rm cn}\, u},
\;\;\;\;\;\;\;\;
{\rm nd}\, u \;=\; \frac{1}{{\rm dn}\, u}.
$$
Quotients are denoted by writing in order the first letters of the numerator and denominator
functions:
\begin{eqnarray*}
&& {\rm sc}\, u \;=\; \frac{{\rm sn}\, u}{{\rm cn}\, u},
\;\;\;\;\;\;\;\;
{\rm sd}\, u \;=\; \frac{{\rm sn}\, u}{{\rm dn}\, u},
\;\;\;\;\;\;\;\;
{\rm cd}\, u \;=\; \frac{{\rm cn}\, u}{{\rm dn}\, u},
\\[5pt]
&&
{\rm cs}\, u \;=\; \frac{{\rm cn}\, u}{{\rm sn}\, u},
\;\;\;\;\;\;\;\;
{\rm ds}\, u \;=\; \frac{{\rm dn}\, u}{{\rm sn}\, u},
\;\;\;\;\;\;\;\;
{\rm dc}\, u \;=\; \frac{{\rm dn}\, u}{{\rm cn}\, u}.
\end{eqnarray*}

The {\it Weierstrass elliptic function} $\wp(z)$ belongs to the first class of elliptic
functions of order 2, those with a single irreducible double pole in each
cell with residue equal to zero. The function $\wp(z)$ is defined by the infinite sum
$$
\wp(z)\;=\;\frac{1}{z^2}+\sum\limits_{(m,n)\neq (0,0)}\left[
\frac{1}{(z-2m\omega_1 - 2n\omega_2)^2} - \frac{1}{(2m\omega_1 - 2n\omega_2)^2}\right],
$$
where $2\omega_1$ and $2\omega_2$ are periods and the summation is taken over all
(positive, negative and zero) integer values of $m$ and $n$, except for when
$m,n$ are both equal to zero.

The function $\wp(z)$ satisfies the differential equation
$
\left(\wp'(z)\right)^2 = 4\wp^{3}(z)-g_2 \wp(z) -g_3,
$
where the {\it elliptic invariants} $g_2$ and $g_3$ are given by
$$
g_2 \;=\; 60 \sum\limits_{(m,n)\neq (0,0)}\frac{1}{(2m\omega_1 + 2n\omega_2)^4},
\;\;\;\;\;\;\;\;
g_3 \;=\; 140 \sum\limits_{(m,n)\neq (0,0)}\frac{1}{(2m\omega_1 + 2n\omega_2)^6}.
$$

Conversely, given
$
\left(dy/dz\right)^2 = 4y^3 - g_2 y - g_3,
$
and if numbers $\omega_1$ and $\omega_2$ can be determined such that
$$
g_2 \;=\; 60 \sum\limits_{(m,n)\neq (0,0)}\frac{1}{(2m\omega_1 + 2n\omega_2)^4},
\;\;\;\;\;\;\;\;
g_3 \;=\; 140 \sum\limits_{(m,n)\neq (0,0)}\frac{1}{(2m\omega_1 + 2n\omega_2)^6}.
$$
then the general solution of the differential equation is
$
y=\wp(z+{\rm const.}).
$

Let
$e_1=\wp(\omega_1)$,
$e_2=\wp(\omega_2)$,
$e_3=\wp(\omega_3)$,
where $\omega_3 = - (\omega_1 +\omega_2)$. The constants $e_1$, $e_2$ and $e_3$
are mutually distinct and are roots of the equation $4y^3-g_2 y - g_3=0$.
It follows that
$$
e_1 + e_2 + e_3 = 0,
\;\;\;\;\;\;\;\;
e_2 e_3 + e_3 e_1 + e_1 e_2 =-\frac{1}{4}\, g_2,
\;\;\;\;\;\;\;\;
e_1 e_2 e_3 = \frac{1}{4}\, g_3.
$$

Let us write
$$
y\;=\; e_3 + \frac{e_1 - e_3}{{\rm sn}^{2}(\lambda z, k)}.
$$
Then, we have
\begin{eqnarray*}
\left(\frac{dy}{dz}\right)^2 &=& 4\lambda^2 (e_1 - e_3)^2\; {\rm ns}^2\,\lambda z
\;{\rm cs}^2\,\lambda z \;{\rm ds}^2\,\lambda z
\\
&=&
4\lambda^2 (e_1 - e_3)^2\; {\rm ns}^2\,\lambda z \left({\rm ns}^2\,\lambda z -1\right)
\left({\rm ns}^2\,\lambda z - k^2\right)
\\
&=&
4\lambda^2 (e_1 - e_3)^{-1}(y-e_3)(y-e_1)\left[y-k^2(e_1 - e_3)-e_3\right].
\end{eqnarray*}
Hence, if $\lambda^2 = e_1 -e_3$ and $k^2=(e_2-e_3)/(e_1-e_3)$, then $y$ satisfies
the equation $(dy/dz)^2 = 4y^3 - g_2 y - g_3$. Therefore,
$$
e_3 + (e_1-e_3)\,{\rm ns}^2\left(z (e_1 - e_3)^{\frac{1}{2}},
\left(\frac{e_2-e_3}{e_1-e_3}\right)^{\frac{1}{2}}\right)
\;=\;\wp(z+h),
$$
where $h$ is a constant. When $z\to 0$, it is seen that $h$ is a period, and so finally one gets
the following relation between Weierstrass and Jacobi elliptic functions:
\begin{equation}\label{PvsSN}
\wp(z)\;=\;e_3 + (e_1-e_3)\,{\rm ns}^2
\left(z (e_1 - e_3)^{\frac{1}{2}},
\left(\frac{e_2-e_3}{e_1-e_3}\right)^{\frac{1}{2}}\right).
\end{equation}
The identity (\ref{PvsSN}) allows to pass from the Jacobian form of the
Lam\'{e} equation (\ref{LameJacobi}) to the Weierstrassian form (\ref{LameW}). Indeed, the change
of independent variable is given by
$
z=\left(u-i{\cal K}'\right)(e_1 - e_3)^{-\frac{1}{2}}.
$
Then, we have\footnote{$k\,{\rm sn}\!\left(\alpha+i{\cal K}',k\right)
={\rm ns}(\alpha, k)$, see \cite{WW} (p. 503, $\S\,22\!\cdot\!34$)}
\begin{eqnarray*}
m\,{\rm sn}^{2}\!\left(u|m\right) &=& m\,{\rm sn}^{2}\!\left(z (e_1 - e_3)^{\frac{1}{2}}+i{\cal K}'|m\right)
\\
&=& {\rm ns}^2\!\left(z (e_1 - e_3)^{\frac{1}{2}}|m\right)
\;=\; \frac{\wp(z)-e_3}{e_1-e_3}.
\end{eqnarray*}
where $m\equiv k^2=(e_2-e_3)/(e_1-e_3)$. Then, the transformed accessory parameter $\mathrm{B}$ in
eq. (\ref{LameW}) is given by
$$
\mathrm{B}\;=\; \mathrm{A}(e_1 - e_3) - \kappa\, e_3 \;\;\Leftrightarrow\;\;
\mathrm{A} \;=\;\frac{\mathrm{B}}{e_1 - e_3} - \frac{1}{3} \kappa (m+1).
$$
\subsection*{Weierstrass $\zeta-$function}
The {\it Weierstrass zeta function} $\zeta(z)$ is defined by the equation
\begin{equation}\label{zeta}
\zeta'(z)=-\wp(z),
\end{equation}
along with the condition $\lim_{z\to\infty}\left(\zeta(z)-\frac{1}{z}\right)=0$.

In terms of the theta function $\theta_1(z|\tau)$ and the constant $\eta_1$ the Weierstrass zeta function
can be expressed as follows
\begin{equation}\label{ZetaFromTheta}
\zeta(z)\;=\;\frac{\partial_{z}\theta_1(z|\tau)}{\theta_1(z|\tau)}+2\eta_1 z.
\end{equation}

\subsection*{Eisenstein series}
The first few {\it Eisenstein series} can be expressed as follows
\begin{eqnarray}\label{E2}
E_{2}(q) &=&
1-24\sum\limits_{n=1}^{\infty}\sigma_{1}(n)q^n=
1-24q - 72q^2 - 96q^3 - 168q^4 - \ldots\;,
\\\label{E4}
E_{4}(q) &=&
1+240\sum\limits_{n=1}^{\infty}\sigma_{3}(n)q^n=
1+240q + 2160q^2 + 6720q^3 + \ldots\;,
\\\label{E6}
E_{6}(q) &=&
1-504\sum\limits_{n=1}^{\infty}\sigma_{5}(n)q^n=
1-504q - 16632q^2 - 122976 q^3 -\ldots\;,
\end{eqnarray}
where $q={\rm e}^{2\pi i \tau}$ and $\sigma_{k}(n)$ is the so-called
{\it divisor function} defined as the sum of the $k$-th powers of
the divisors of $n$, $\sigma_{k}(n)=\sum_{d|n} d^{k}$.

\subsection*{Dedekind $\eta-$function}
$$
\eta(q) \;=\; q^{\frac{1}{24}}\prod_{n=1}^{\infty}(1-q^{n}),
\;\;\;\;\;\;\;\;\;\;\;
q={\rm e}^{2\pi i \tau}.
$$

\section{Integrals ${\k K}_n$}
\renewcommand{\theequation}{B.\arabic{equation}}
\setcounter{equation}{0}
The integral
$$
{\k K}_1\;=\;\oint\limits_{\cal A} \wp(z) dz
$$
one can compute directly using (\ref{zeta}) and (\ref{ZetaFromTheta}).
The result is
$
{\k K}_1=-2\eta_1.
$
In order to compute the integrals
$$
{\k K}_n\;=\;\oint\limits_{\cal A} \wp(z)^{n} dz,
\;\;\;\;\;\;\;\;\;\;n>1
$$
one can employ the following relation between powers of $\wp$ and its even
derivatives (see \cite{GV} and refs. therein):
\begin{equation}\label{wprelation}
\wp^n \;=\; B_{n}^{(n)}+\sum\limits_{r=0}^{n-1}\frac{B^{(n)}_{r}}{\left(2n-2r-1\right)!}\,
\wp^{(2n-2-2r)}.
\end{equation}
Quantities $B^{(n)}_{r}$ above --- the so-called {\it Halphen coefficients} --- are given
by the recurrence relation:
\begin{eqnarray*}
B^{(n)}_{r}\;=\;\frac{(2n-2r-2)(2n-2r-1)}{(2n-2)(2n-1)}\,B^{(n-1)}_{r}
+
\frac{2n-3}{4(2n-1)}\,B^{(n-2)}_{r-2}\,g_2 + \frac{n-2}{2(2n-1)}\,B^{(n-3)}_{r-3}\,g_3
\end{eqnarray*}
with $n>0$, $r=0,\ldots,n$; $B^{(n)}_{r}=0$ for $r<0$ or $r>n$; $B^{(n)}_{0}=1$ and
$B^{(n)}_{1}=0$ for any $n$. $B^{(n)}_{r}$'s are then polynomials in $g_2$, $g_3$ with
rational positive coefficients.

For $n=2,3,4$  the eq. (\ref{wprelation}) yields
\begin{eqnarray*}
\wp^2 &=& B^{(2)}_{2}+\frac{1}{6}B^{(2)}_{0}\wp^{(2)}+B^{(2)}_{1}\wp
=
\frac{1}{12}g_2 + \frac{1}{6}\wp^{(2)},
\\
\wp^3 &=& B^{(3)}_{3}+\frac{1}{120}B_{0}^{(3)}\wp^{(4)}+\frac{1}{6}B_{1}^{(3)}\wp^{(2)}
+B^{(3)}_{2}\wp
=
\frac{1}{10}g_3 + \frac{1}{120}\wp^{(4)}+\frac{3}{20}g_2 \wp,
\\
\wp^4 &=& B^{(4)}_{4} + \frac{1}{5040}B^{(4)}_{0}\wp^{(6)}
+\frac{1}{120}B^{(4)}_{1}\wp^{(4)}
+\frac{1}{6}B^{(4)}_{2}\wp^{(2)}+B^{(4)}_{3}\wp
\\
&=& \frac{5}{336}g^{2}_{2}+\frac{1}{5040}\wp^{(6)}+\frac{1}{30}g_2 \wp^{(2)}+
\frac{1}{7}g_3 \wp.
\end{eqnarray*}
Then,
\begin{eqnarray*}
{\k K}_{2} &=& \frac{1}{12}g_2,
\\
{\k K}_{3} &=& \frac{1}{10}g_3 + \frac{3}{20}g_2{\k K}_{1} =
\frac{1}{10}g_3 - \frac{3}{10}g_2{\eta}_{1},
\\
{\k K}_{4} &=& \frac{5}{336}g^{2}_{2}+\frac{1}{7}g_3 {\k K}_{1}
=\frac{5}{336}g^{2}_{2}-\frac{2}{7}g_3 {\eta}_{1}.
\end{eqnarray*}
Now, in order to get the coefficients (\ref{e1})--(\ref{e4}) of the expansion of the energy eigenvalue
one has to use the following relations:
$$
\eta_1\;=\;\frac{4\pi^2}{24}\,E_2,
\;\;\;\;\;\;\;\;
g_2\;=\;\frac{4\pi^4}{3}\,E_4,
\;\;\;\;\;\;\;\;
g_3\;=\;\frac{8\pi^6}{27}\,E_6.
$$

\section{Coefficients of the torus quantum conformal block}
\begin{eqnarray*}
{\cal F}_{c,\Delta}^{\Delta_\beta,1}&=&
\frac{\left(\Delta_\beta-1\right)\Delta_\beta}{2\Delta }+1,
\\[10pt]
{\cal F}_{c,\Delta}^{\Delta_\beta,2}&=&
\left[4 \Delta
\left(2 c \Delta +c+16 \Delta ^2-10 \Delta \right)\right]^{-1}
\\
&&
\Big[\left(8 c \Delta +3 c+128 \Delta ^2+56 \Delta \right)
   \Delta _{\beta }^2+\left(-8 c \Delta -2 c-128 \Delta
   ^2\right) \Delta _{\beta }
\\
&+&
   (c+8 \Delta ) \Delta _{\beta
   }^4+(-2 c-64 \Delta ) \Delta _{\beta }^3+16 c \Delta
   ^2+8 c \Delta +128 \Delta ^3-80 \Delta ^2\Big].
\\
\ldots\;&&.
\end{eqnarray*}

\acknowledgments
I'm grateful to Franco Ferrari for useful discussions, very valuable advices
and his kind hospitality during my stays in Szczecin.
I'm also grateful to Zbigniew Jask\'{o}lski and Artur Pietrykowski
for stimulating questions and comments.

This research has been supported in part by the Polish National
Science Centre under Grant No. N202 326240.


\begin{thebibliography}{99}

\bibitem{WW}
E.T. Whittaker, G.N. Watson, {\it A course of modern analysis}, Cambridge Univ. Press (1952).


\bibitem{Maier}
R.S. Maier, {\it Lam\'{e} polynomials, hyperelliptic reductions and Lam\'{e} band structure},
Philos. Trans. Roy. Soc. London Ser. A 366 (2008), 1115-1153, math-ph/0309005.


\bibitem{Ply1}
F. Correa, L.-M. Nieto, M.S. Plyushchay, {\it Hidden nonlinear supersymmetry of finite-gap Lame equation},
Phys. Lett. B644 (2007) 94-98, hep-th/0608096.


\bibitem{Ply2}
F. Correa, M.S. Plyushchay,
{\it Peculiarities of the hidden nonlinear supersymmetry of 
Poschl-Teller system in the light of Lame equation},
J. Phys. A40 (2007) 14403-14412, arXiv:0706.1114 [hep-th].


\bibitem{Ply3}
F. Correa, V. Jakubsky, L.-M. Nieto, M.S. Plyushchay,	
{\it Self-isospectrality, special supersymmetry, and their effect on the band structure},
Phys. Rev. Lett. 101 (2008) 030403, arXiv:0801.1671 [hep-th].


\bibitem{Ply4}
F. Correa, V. Jakubsky, M.S. Plyushchay,
{\it Finite-gap systems, tri-supersymmetry and self-isospectrality},
J. Phys. A41 (2008) 485303, arXiv:0806.1614 [hep-th].


\bibitem{Ply5}
F. Correa, G.V. Dunne, M.S. Plyushchay,
{\it The Bogoliubov/de Gennes system, the AKNS hierarchy, and nonlinear quantum mechanical supersymmetry},
Annals Phys. 324 (2009) 2522-2547, arXiv:0904.2768 [hep-th].


\bibitem{Ply6}
M.S. Plyushchay, A. Arancibia, L.-M. Nieto,
{\it Exotic supersymmetry of the kink-antikink crystal, and the infinite period limit},
Phys. Rev. D83 (2011) 065025, arXiv:1012.4529 [hep-th].


\bibitem{Ply7}
A. Arancibia, M.S. Plyushchay,
{\it Extended supersymmetry of the self-isospectral crystalline and soliton chains},
Phys. Rev. D85 (2012) 045018, arXiv:1111.0600 [hep-th].


\bibitem{NekraSha}
N. Nekrasov, S. Shatashvili, {\it Quantization of Integrable Systems
and Four Dimensional Gauge Theories}, hep-th/0908.4052.


\bibitem{NekraSha2}
N. Nekrasov, S. Shatashvili, {\it Supersymmetric vacua and Bethe
ansatz}, In ``Cargese 2008, Theory and Particle Physics: the LHC
perspective and beyond'', hep-th/0901.4744.


\bibitem{NekraSha3}
N. Nekrasov, S. Shatashvili, {\it Quantum integrability and
supersymmetric vacua}, Prog. Theor. Phys. Suppl. 177 (2009) 105-119, hep-th/0901.4748.


\bibitem{Iachello}
Y. Alhassid,  F. G�rsey, F. Iachello,
{\it Potential scattering, transfer matrix and group theory}, Phys. Rev. Lett. 50, 873 (1983).


\bibitem{Caputo}
J.-G. Caputo, N. Flytzanis, Y. Gaididei, N. Stefanakis, E. Vavalis,
{\it Stability analysis of static solutions in a Josephson junction}, Supercond. Sci. Technol. 13 (2000),
423-438, cond-mat/0010335.


\bibitem{Maier2}
R.S. Maier, L. D. Stein, {\it Droplet nucleation and domain wall motion in a bounded
interval}, Phys. Rev. Lett. 87 (2001) 270601, cond-mat/0108217.


\bibitem{Kantowski}
R. Kantowski,  R.C. Thomas, {\it Distance-Redshift in Inhomogeneous $\Omega_0 = 1$
Friedmann-Lemaitre-Robertson-Walker Cosmology}, Astrophys. J. 561 (2001) 491-495,
astro-ph/0011176.


\bibitem{Boyanovski}
D. Boyanovsky, H.J. de Vega, R. Holman, J.F.J. Salgado,
{\it Analytic and numerical study of preheating dynamics}, Phys. Rev. D54 (1996) 7570-7598,
hep-ph/9608205.


\bibitem{Greene}
P. Greene, L. Kofman, A. Linde, A. Starobinsky,
{\it Structure of Resonance in Preheating after Inflation}, Phys. Rev. D 56 (1997) 6175-6192,
hep-ph/9705347.


\bibitem{Kaiser}
D.I. Kaiser, {\it Resonance structures for preheating with massless fields},
Phys. Rev. D 57 (1998) 702-711, hep-ph/9707516.


\bibitem{Ivanov}
P. Ivanov, {\it On Lam\'{e}'s equation of a particular kind}, J. Phys. A34 (2001) 8145-8150,
math-ph/0008008.


\bibitem{M-K}
H.J.W. M\"{u}ller-Kirsten,
{\it Introduction to Quantum Mechanics: Schr\"{o}dinger Equation and Path Integral},
World Scientific, Singapore, 2006.


\bibitem{Finkel}
F. Finkel, A. Gonzalez-Lopez, M.A. Rodriguez,
{\it A New Algebraization of the Lame Equation}, J. Phys. A: Math. Gen. 33 (2000) 1519-1542,
math-ph/9908002.


\bibitem{KRV}
L. Keen, H.E. Rauch, A.T. Vasquez,
{\it Moduli of punctured tori and the accessory parameter of Lam\'{e}'s equation},
Trans. Am. Math. Soc. 255 (1979).


\bibitem{BMT}
G. Bonelli, K. Maruyoshi, A. Tanzini,
{\it Quantum Hitchin Systems via beta-deformed Matrix Models}, arXiv:1104.4016 [hep-th].

\bibitem{Bernard}
D. Bernard, {\it On The Wess-Zumino-Witten Models On The Torus}, Nucl. Phys. B 303, 77 (1988).


\bibitem{Etingof}
P. I. Etingof and A. A. Kirillov,
{\it Representation of affine Lie algebras, parabolic differential equations and Lame functions},
arXiv:hep-th/9310083.


\bibitem{FW}
G. Felder, C. Weiczerkowski,
{\it Conformal blocks on elliptic curves and the Knizhnik-Zamolodchikov-Bernard equations},
Commun. Math. Phys. 176, 133-162 (1996), [hep-th/9411004].


\bibitem{AT}
L. F. Alday and Y. Tachikawa,
{\it Affine SL(2) conformal blocks from 4d gauge theories},
Lett. Math. Phys. 94, 87-114 (2010), [arXiv:1005.4469 [hep-th]].

\bibitem{Menotti1}
P. Menotti, {\it Accessory parameters for Liouville theory on the torus},
JHEP 12 (2012) 001, arXiv:1207.6884 [hep-th].


\bibitem{Menotti2}
P. Menotti, {\it Riemann-Hilbert treatment of Liouville theory on the torus},
J. Phys. A 44 115403, (2011), hep-th/1010.4946.


\bibitem{Menotti3}
P. Menotti, {\it Riemann-Hilbert treatment of Liouville theory on the torus: The general case},
J. Phys. A 44 335401, (2011), hep-th/1104.3210.


\bibitem{Menotti:2013bka}
P. Menotti, {\it Hyperbolic deformation of the strip-equation and the accessory parameters for the torus},
arXiv:1307.0306 [hep-th].


\bibitem{Zamolodchikov:1995aa}
A.~B.~Zamolodchikov and A.~B.~Zamolodchikov, {\it Structure
constants and conformal bootstrap in Liouville field theory}, Nucl.\
Phys.\ B 477 (1996) 577, hep-th/9506136.


\bibitem{ZamClassBlock}
A. Litvinov, S. Lukyanov, N. Nekrasov, A. Zamolodchikov,
{\it Classical Conformal Blocks and Painleve VI}, arXiv:1309.4700 [hep-th].


\bibitem{NekraRosSha}
N. Nekrasov, A. Rosly, S. Shatashvili, {\it Darboux coordinates, Yang-Yang functional, and gauge theory},
Nucl. Phys. Proc. Suppl. 216, (2011),  69-93,
hep-th/1103.3919.


\bibitem{Teschner}
J. Teschner, {\it Quantization of the Hitchin moduli spaces,
Liouville theory, and the geometric Langlands correspondence}, hep-th/1005.2846.


\bibitem{Piatek}
M. Pi\c{a}tek, {\it Classical conformal blocks from TBA for the elliptic Calogero-Moser system},
JHEP 06 (2011) 050, hep-th/1102.5403.


\bibitem{FFPiatek}
F. Ferrari and M. Pi\c{a}tek, {\it Liouville theory, ${\cal N}=2$
  gauge theories and accessory parameters}, JHEP 05 (2012) 025, arXiv:1202.2149 [hep-th].


\bibitem{Hartman}
T. Hartman, {\it Entanglement Entropy at Large Central Charge}, arXiv:1303.6955 [hep-th].


\bibitem{Pogho1}
R. Poghossian, {\it Deforming SW curve}, JHEP 04 (2011) 033, hep-th/1006.4822.


\bibitem{Belavin:1984vu}
A.~A.~Belavin, A.~M.~Polyakov and A.~B.~Zamolodchikov, {\it Infinite
Conformal Symmetry In Two-Dimensional Quantum Field Theory}, Nucl.\
Phys.\ B 241, (1984) 333.


\bibitem{EO}
T. Eguchi, H. Ooguri, {\it Conformal and Current Algebras on General
Riemann Surface}, Nucl. Phys. B 282 (1987) 308�328.


\bibitem{JT0}
J. Teschner, {\it An analog of a modular functor from quantized Teichm\"{u}ller theory},
math/0510174.


\bibitem{Zam}
A.~B.~Zamolodchikov, {\it Conformal symmetry in two-dimensional
space: recursion representation of conformal block},
Theor. Math. Phys. 73 (1987) 1088.


\bibitem{Zamolodchikov:ie}
A.~B.~Zamolodchikov, {\it Conformal Symmetry In Two-Dimensions: An
Explicit Recurrence Formula For The Conformal Partial Wave
Amplitude}, Commun.\ Math.\ Phys.\ 96 (1984) 419.


\bibitem{FL}
V.A. Fateev, A.V. Litvinov, {\it On AGT conjecture}, JHEP 02 (2010) 014,
hep-th/09120504.


\bibitem{Pogho}
R. Poghossian, {\it Recursion relations in CFT and N=2 SYM theory}, JHEP 12 (2009) 038,
hep-th/0909.3412.


\bibitem{HJStorus}
L. Hadasz, Z. Jask\'{o}lski, P. Suchanek, {\it Recursive
representation of the torus 1-point block}, JHEP 01 (2010) 063, 
hep-th/0911.2353.


\bibitem{Zam0}
A.~B.~Zamolodchikov, {\it Two-dimensional conformal symmetry and
critical four-spin correlation functions in the Ashkin-Teller
model}, Sov. Phys. JEPT 63 (5) (1986) 1061.


\bibitem{HJP}
L. Hadasz, Z. Jask\'{o}lski, M. Pi\c{a}tek, {\it Classical geometry
from the quantum Liouvill theory}, Nucl. Phys. B 724 529 (2005), hep-th/0504204.


\bibitem{AGT}
L. Alday, D. Gaiotto, Y. Tachikawa, {\it Liouville Correlation
Functions from Four-dimensional Gauge Theories}, Lett. Math. Phys.
91 (2010) 167-197, hep-th/0906.3219.


\bibitem{Wyllard}
N. Wyllard, {\it $A_{N-1}$ conformal Toda field theory correlation
functions from conformal N = 2 SU(N) quiver gauge theories}, JHEP 11 (2009) 002, hep-th/0907.2189.


\bibitem{MMU(3)}
A. Mironov, A. Morozov, {\it On AGT relation in the case of U(3)},
Nucl. Phys. B 825 (2010) 1-37, hep-th/0908.2569.


\bibitem{MMMM}
A. Mironov, S. Mironov, A. Morozov, A. Morozov, {\it CFT exercises
for the needs of AGT}, hep-th/0908.2064.


\bibitem{MM2}
A. Mironov, A. Morozov, {\it Proving AGT relations in the large-c
limit}, Phys. Lett. B 682 (2009) 118-124, hep-th/0909.3531.


\bibitem{Hadasz:2010xp}
L.~Hadasz, Z.~Jaskolski and P.~Suchanek,
{\it Proving the AGT relation for $N_f = 0,1,2$ antifundamentals}, JHEP 1006 (2010) 046,
arXiv:1004.1841 [hep-th].


\bibitem{Alba:2010qc}
V. A. Alba, V. A. Fateev, A. V. Litvinov, G. M. Tarnopolskiy,	
{\it On combinatorial expansion of the conformal blocks arising from AGT conjecture},
Lett. Math. Phys. 98 (2011) 33-64, arXiv:1012.1312 [hep-th].


\bibitem{Vartanov:2013ima}
  G.~Vartanov and J.~Teschner,
  {\it Supersymmetric gauge theories, quantization of moduli spaces of flat connections, and conformal field theory},
  arXiv:1302.3778 [hep-th].


\bibitem{SV}
O. Schiffmann, E. Vasserot,
{\it Cherednik algebras, W algebras and the equivariant cohomology
of the moduli space of instantons on $\mathbb{A}^2$}, arXiv:1202.2756.


\bibitem{MO}
D. Maulik, A. Okounkov, {\it Quantum Groups and Quantum Cohomology}, arXiv:1211.1287.


\bibitem{MCTan}
M.-C.Tan, {\it M-Theoretic Derivations of 4d-2d Dualities:
From a Geometric Langlands Duality for Surfaces, to the AGT Correspondence, to Integrable Systems},
JHEP 07 (2013) 171, arXiv:1301.1977 [hep-th].


\bibitem{Matsuo1}
S. Kanno, Y. Matsuo, H. Zhang, {\it Virasoro constraint for Nekrasov instanton partition function},
JHEP 10 (2012) 097, arXiv:1207.5658 [hep-th].


\bibitem{Matsuo2}
S. Kanno, Y. Matsuo, H. Zhang,
{\it Extended Conformal Symmetry and Recursion Formulae for Nekrasov Partition Function},
JHEP 08 (2013) 028, arXiv:1306.1523 [hep-th].


\bibitem{N}
N. Nekrasov, {\it Seiberg-Witten  prepotential  from instanton
counting}, Adv. Theor. Math. Phys. 7 (2004) 831- 864, hep-th/0206161.


\bibitem{NekraOkun}
N. Nekrasov, A. Okounkov, {\it Seiberg-Witten theory and random
partitions}, hep-th/0306238.


\bibitem{SW1}
N. Seiberg and E. Witten,
{\it Monopole Condensation, And Confinement In N=2 Supersymmetric Yang-Mills Theory},
Nucl. Phys. B 426 (1994) 19-52, arXiv:9407087 [hep-th].


\bibitem{SW2}
N. Seiberg and E. Witten, {\it Monopoles,  duality and chiral symmetry breaking in N = 2
supersymmetric QCD}, Nucl. Phys. B 431 (1994) 484, arXiv:9408099 [hep-th].


%
%
%
%


\bibitem{YY}
C.  N.  Yang,  C.  P.  Yang, {\it Thermodinamics  of  a
one-dimensional  system  of  bosons  with  repulsive delta-function
interaction}, J. Math. Phys. 10 (1969) 1115.


\bibitem{Fucito}
F. Fucito, J.F. Morales, D.R. Pacifici, R. Poghossian, {\it Gauge theories on $\Omega$-backgrounds from
non commutative Seiberg-Witten curves}, JHEP  05  (2011)   098,  hep-th/1103.4495.


\bibitem{Picard1}
    E.~Picard,
    {\it De l'\'{e}quation $\Delta_2 u = k{\rm e}^u$ sur une surface de Riemann ferm\'{e}e }
    J.\ Math.\ Pure Appl. (4) 9 (1893), 273 - 291.


\bibitem{Picard2}
    E.~Picard,
    {\it De l'int\'{e}egration de l'\'{e}quation $\Delta u = {\rm e}^u$ sur une surface de Riemann ferm\'{e}e }
    Crelle's J. 130 (4) 9 (1905), 243 - 258.


\bibitem{Troyanov}
    M.~Troyanov,
    {\it Prescribing curvature on compact surfaces with conical singularities }
    Trans.\ Amer.\ Math.\ Soc. 134 (1991), 793 - 821.


\bibitem{Heins}
M. Heins, {\it On a class of conformal metrics}, Nagoya Math.J. 21 (1962) 1  60.


\bibitem{Tak}
L. A. Takhtajan, {\it Topics  in  quantum  geometry  of  Riemann
surfaces: Two-dimensional quantum gravity}, published  in  Como  Quantum  Groups  1994:541-580, hep-th/9409088.


\bibitem{TZ2}
P. G. Zograf, L. A. Takhtajan, {\it On Liouville equation, accessory
parameters and the geometry of Teichm\"{u}ller space for Riemann surface of genus  0},
Math. USSR Sbornik 60 (1988) 143.


\bibitem{ZoTa2}
    P.~G.~Zograf and L.~A.~Takhtajan,
    {\it On uiformization of Riemann surfaces and the Weil-Petersson metric
    on Teichm\"{u}ller and Schottky spaces.}
    Math. USSR Sbornik 60, (1988) 297.


\bibitem{TZ}
L. Takhtajan, P. Zograf, {\it Hyperbolic  2-spheres  with  conical
singularities,  accessory parameters and K\"{a}hler metrics on $M_{0,n}$},
Trans. Amer. Math. Soc. 355 (2003), 1857-1867, math.cv/0112170.


\bibitem{Cantini:2001wr}
    L.~Cantini, P.~Menotti and D.~Seminara,
    {\it Proof of Polyakov conjecture for general elliptic singularities,}
    Phys.\ Lett. B 517  (2001) 203, hep-th/0105081.


\bibitem{Hadasz:2003kp}
    L.~Hadasz and Z.~Jaskolski,
    {\sl Polyakov conjecture for hyperbolic singularities},
    Phys.\ Lett. B 574 (2003) 129, hep-th/0308131.


\bibitem{sei}
    N.~Seiberg,
    {\it Notes on Quantum Liouville Theory and Quantum Gravity}
    in
    {\em Common Trends in Mathematics and Quantum
    Field Theory},
    Proc. of the 1990 Yukawa International Seminar,
    Prog.\ Theor.\ Phys.\ Suppl.\  102 (1990) 319.


\bibitem{Hadasz:2003he}
    L.~Hadasz and Z.~Jaskolski,
    {\it Classical Liouville action on the sphere with three hyperbolic singularities},
    Nucl.\ Phys.\ B 694, 493 (2004),
    arXiv:hep-th/0309267.



\bibitem{JT1}
J. Teschner, {\it Liouville theory revisited}, Class. Quant. Grav. 18 (2001),
hep-th/0104158.




\bibitem{DO}
H. Dorn and H. Otto, {\it Two and three point functions in Liouville theory},
Nucl. Phys. B 429 (1994) 375, hep-th/9403141.




\bibitem{Sonoda}
H. Sonoda, {\it Sewing Conformal Field Theories. 2}, Nucl. Phys. B 311 (1988) 417.


\bibitem{HJSmodular}
L. Hadasz, Z. Jaskolski, P. Suchanek,
{\it Modular bootstrap in Liouville field theory}, Phys. Lett. B 685 (2010) 79-85,
arXiv:0911.4296 [hep-th].




\bibitem{T}
J. Teschner, {\it On the Liouville three point function}, Phys. Lett. B 363
(1995) 65-70, hep-th/9507109.




\bibitem{Nakayama}
Y.~Nakayama, {\it Liouville Field Theory --- A decade after revolution},
Int. J. Mod. Phys. A 19 (2004) 2771-2930, hep-th/0402009.






\bibitem{Witten}
D. Harlow,  J. Maltz, E. Witten,
{\it Analytic Continuation of Liouville Theory},
JHEP 12 (2011) 071, 
arXiv:1108.4417 [hep-th].


\bibitem{KashaniPoor:2012wb}
  A.~-K.~Kashani-Poor and J.~Troost,
  {\it The toroidal block and the genus expansion},
  JHEP 1303 (2013) 133,
  arXiv:1212.0722 [hep-th].

\bibitem{MirMor}
A.~Mironov, A.~Morozov, {\it Nekrasov Functions and Exact Bohr-Sommerfeld Integrals},
JHEP 04 (2010) 040, arXiv:0910.5670 [hep-th].




\bibitem{Maruyoshi:2010iu}
  K.~Maruyoshi and M.~Taki,
  {\it Deformed Prepotential, Quantum Integrable System and Liouville Field Theory},
  Nucl.\ Phys.\ B 841 (2010) 388,
  arXiv:1006.4505 [hep-th].


\bibitem{He:2011zk}
  W.~He, {\it Combinatorial approach to Mathieu and Lame equations},
  arXiv:1108.0300 [math-ph].




\bibitem{MK}
H.~J.~W. M\"{u}ller-Kirsten, {\it Introduction to quantum mechanics:
Schr\"{o}dinger equation and path integral},  World Scientific, Singapore, 2006.


\bibitem{Dunne:1999zc}
G.~V.~Dunne and K.~Rao, {\it Lame instantons},
JHEP 0001 (2000) 019, hep-th/9906113.


\bibitem{Langmann:2004sj}
  E.~Langmann,
  {\it An Explicit solution of the (quantum) elliptic Calogero-Sutherland model},
  math-ph/0407050.


\bibitem{Langmann}
E. Langmann, {\it Explicit solution of the (quantum) elliptic Calogero-Sutherland model}, math-ph/0401029.


\bibitem{Billo1}
M. Billo, M. Frau, L. Gallot, A. Lerda, I. Pesando,
{\it Deformed N=2 theories, generalized recursion relations and S-duality}, JHEP 04 (2013) 039,
arXiv:1302.0686 [hep-th].


\bibitem{Billo2}
M. Billo, M. Frau, L. Gallot, A. Lerda, I. Pesando,
{\it Modular anomaly equation, heat kernel and S-duality in N=2 theories},
arXiv:1307.6648 [hep-th].

\bibitem{DS}
Vl.S. Dotsenko, V.A. Fateev, 
{\it Conformal algebra and multipoint correlation functions in 2D statistical models}, 
Nucl. Phys. B 240 (1984) 312-348.


\bibitem{MMS2}	
A. Mironov, A. Morozov, Sh. Shakirov
{\it Conformal blocks as Dotsenko-Fateev Integral Discriminants}, Int. J. Mod. Phys. A25 (2010) 3173-3207,
arXiv:1001.0563 [hep-th].


\bibitem{MMS3}	
A. Mironov, A. Morozov, Sh. Shakirov,
{\it On 'Dotsenko-Fateev' representation of the toric conformal blocks}
J. Phys. A 44 (2011) 085401, arXiv:1010.1734 [hep-th].


\bibitem{MMM2}
A. Mironov,  Al. Morozov and And. Morozov,
{\it Matrix model version of AGT conjecture and generalized Selberg integrals},
Nucl. Phys. B 843 (2011) 534-557, \tt arXiv:1003.5752 [hep-th]\rm.



\bibitem{Dijkgraaf} R. Dijkgraaf and C. Vafa, {\it Toda Theories,
Matrix Models, Topological Strings and N = 2
Gauge Systems}, \tt arXiv:0909.2453 [hep-th]\rm.


\bibitem{Sulkowski}
P. Sulkowski, {\it  Matrix models for beta-ensembles from
Nekrasov partition functions}, JHEP 04 (2010) 063, 
\tt arXiv:0912.5476 [hep-th]\rm.


\bibitem{Bourgine1}
J.-E. Bourgine,
{\it Notes on Mayer Expansions and Matrix Models},
arXiv:1310.3566 [hep-th].


\bibitem{Bourgine2}
J.-E. Bourgine,
{\it Large N techniques for Nekrasov partition functions and AGT conjecture},
JHEP 05 (2013) 047, arXiv:1212.4972 [hep-th].


\bibitem{Bourgine3}
J.-E. Bourgine, {\it Large N limit of beta-ensembles and deformed Seiberg-Witten relations}
JHEP 08 (2012) 046, arXiv:1206.1696.


\bibitem{MMM}
A. Marshakov, A. Mironov, A. Morozov,
{\it On AGT Relations with Surface Operator Insertion and Stationary Limit of Beta-Ensembles},
J. Geom. Phys. 61  (2011) 1203-1222,
\tt arXiv:1011.4491 [hep-th]\rm.


\bibitem{MMS}
A. Mironov, A. Morozov, Sh. Shakirov,
{\it Matrix Model Conjecture for Exact BS Periods and Nekrasov Functions},
JHEP 02 (2010) 030, \tt arXiv:0911.5721 [hep-th] \rm.


\bibitem{Morozov}
A. Morozov, {\it Challenges of beta-deformation},
\tt arXiv:1201.4595 [hep-th]\rm.


\bibitem{FPplb}
F. Ferrari, M. Piatek,
{\it On a singular Fredholm-type integral equation arising in N=2 super Yang-Mills theories}
Phys. Lett. B 718 (2013) 1142-1147, arXiv:1202.5135 [hep-th].


\bibitem{FPcjp}
F. Ferrari, M. Piatek,
{\it On a path integral representation of the Nekrasov
instanton partition function and its Nekrasov-Shatashvili limit}, arXiv:1212.6787.


\bibitem{GV}
M. Grosset, A.P. Veselov, {\it Elliptic Faulhaber polynomials and Lam\'{e} densities of states}, math-ph/0508066.
\end{thebibliography}
\end{document}